\documentclass[
amsfonts,
amsmath,
aps,
prd,
nofootinbib,
preprint,
superscriptaddress
]{revtex4}
\usepackage[dvipdfmx]{graphicx}
\usepackage{bm,latexsym,amsmath,amssymb,amsfonts} 
\DeclareMathAlphabet{\mathpzc}{OT1}{pzc}{m}{it}
\usepackage[normalem]{ulem}
\usepackage[breaklinks=true,colorlinks=true]{hyperref}
\usepackage{xcolor}
\usepackage{wasysym}
\usepackage[mathscr]{eucal}
\usepackage{multirow}
\usepackage{enumitem}
\usepackage{tabularray}
\usepackage{tablefootnote}
\usepackage{longtable}
\usepackage{booktabs}
\usepackage{lipsum}

\newcommand{\orcid}[1]{\href{https://orcid.org/#1}{\includegraphics[width=8pt]{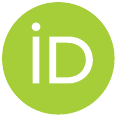}}}
\definecolor{royalblue}{rgb}{0.0, 0.0, 0.8}
\hypersetup{linkcolor=royalblue, citecolor=royalblue, urlcolor=royalblue}

\begin{document}

\def\Journal#1#2#3{\href{https://doi.org/#3}{{#1} #2}}
\def\arXiv#1#2{\href{https://arxiv.org/abs/#1}{arXiv:#1}}
\def\supplmat{\hyperref[sec. supplementary material]{Supplementary Material}}
        
\title{Photon propagation and black hole Imaging in Kruglov nonlinear electrodynamics}

\author{H.~S.~Ramadhan\orcid{0000-0003-0727-738X}}
\email{hramad@sci.ui.ac.id}
\thanks{Corresponding author.}
\affiliation{Departemen Fisika, FMIPA, Universitas Indonesia, Depok 16424, Indonesia}

\author{M.~F.~Fauzi\orcid{0009-0005-3380-6005}}
\email{muhammad.fahmi31@ui.ac.id}
\affiliation{Departemen Fisika, FMIPA, Universitas Indonesia, Depok 16424, Indonesia}

\author{D.~A.~Witjaksana}
\email{daniel.witjaksana@student.kuleuven.be}
\affiliation{Department of Physics and Astronomy, KU Leuven, Celestijnenlaan 200D, B-3001 Leuven, Belgium.}

\author{A.~Sulaksono\orcid{0000-0002-1493-5013}}
\email{anto.sulaksono@sci.ui.ac.id}
\affiliation{Departemen Fisika, FMIPA, Universitas Indonesia, Depok 16424, Indonesia}

\begin{abstract}

Nonlinear electrodynamics (NLED) modifies photon propagation through an effective optical geometry, allowing optical phenomena to deviate from those predicted solely by the underlying spacetime metric. In this work, we revisit the Kruglov model, which generalizes Born--Infeld electrodynamics, and investigate the corresponding null geodesics in theeffective geometry through direct numerical integration, without relying on strong-field analytical approximations. We analyze the resulting photon spheres, gravitational lensing, black-hole shadows, and thin accretion-disk images, together with their dependence on the parameter $q$. For sufficiently small positive $q$, the effective angular metric develops a pole that generates stable photon orbits outside the event horizon and locally reverses the azimuthal evolution of photon trajectories, producing a distinctive bending pattern that resembles, although does not originate from, frame dragging. Conversely, negative $q$ introduces a zero of the effective correction factor, which acts as a turning point for photon trajectories and prevents them from entering the region where the background metric becomes complex. It produces measurable changes in light deflection, shadow radius, photon-ring morphology, and relativistic images. In particular, smaller positive values of $q$ produce thinner photon rings and enlarged shadow radii, whereas negative values generally lead to broader photon rings and smaller shadows. Comparison with the current Event Horizon Telescope constraints on Sgr~A* shows that a substantial region of the Kruglov parameter space remains compatible with present observations. 

\end{abstract}

\maketitle
\newpage
\section{Introduction}
\label{intro}

Gravitational light deflection is one of the classical and most profound predictions of General Relativity (GR). Using GR, Einstein showed that light passing near a massive body experiences a deflection angle twice the Newtonian prediction based on the equivalence principle, a result later confirmed by the 1919 Eddington expedition~\cite{Schneider:1992bmb, Schneider2, DE1919}. The term ``lens’’ to describe gravitational light deflection was first introduced by Lodge~\cite{Lodge}, while the modern framework of gravitational lensing was developed through the pioneering works of Liebes~\cite{Liebes:1964zz} and Refsdal~\cite{Refsdal:1964yk} (see~\cite{Schneider:1992bmb, Schneider2, Blandford:1991xc, Perlick:2004tq}). In the context of black hole (BH), Darwin~\cite{darwin, Bozza2010} was the first to demonstrate that the deflection angle in the Schwarzschild spacetime exhibits a logarithmic divergence in the strong-field regime, establishing the basis of strong gravitational lensing by BHs.


The presence of an event horizon together with strong light bending gives rise to the phenomenon known as the BH \textit{shadow}~\cite{Luminet:1979nyg, Falcke:1999pj}. Light rays with impact parameters below a critical value are captured by the BH and fail to reach a distant observer, resulting in a dark region in the observer’s sky corresponding to the apparent projection of the photon capture region. Observationally, the Event Horizon Telescope (EHT) collaboration has successfully imaged the shadows of the supermassive BHs in M87*~\cite{EventHorizonTelescope:2019dse,EventHorizonTelescope:2019uob,EventHorizonTelescope:2019jan,EventHorizonTelescope:2019ths,EventHorizonTelescope:2019pgp,EventHorizonTelescope:2019ggy,EventHorizonTelescope:2021bee,EventHorizonTelescope:2021srq,EventHorizonTelescope:2023gtd} and Sagittarius A* (Sgr~A*)~\cite{EventHorizonTelescope:2022wkp,EventHorizonTelescope:2022apq,EventHorizonTelescope:2022wok,EventHorizonTelescope:2022exc,EventHorizonTelescope:2022urf,EventHorizonTelescope:2022xqj,EventHorizonTelescope:2024hpu,EventHorizonTelescope:2024rju}. The reconstructed images reveal a bright emission ring surrounding a central dark region, in qualitative agreement with general relativistic simulations.

These developments have motivated extensive studies of strong-field optical phenomena as probes of BH spacetimes and alternative compact-object models. In particular, ray-tracing calculations of gravitational lensing, photon rings, and accretion-disk images have become important tools for connecting theoretical models with horizon-scale observations. Recent studies have further emphasized that observable image features can depend not only on the background spacetime geometry, but also on the structure of photon propagation itself and the properties of the emitting source~\cite{Boos:2025nzc}. This motivates the investigation of theories beyond Maxwell electrodynamics in which photon trajectories deviate from the standard null geodesics of the underlying spacetime.


One possible mechanism leading to such modifications arises from nonlinear electrodynamics (NLED). The idea of NLED itself is well-established in theoretical physics. As early as 1934, Born and Infeld proposed a modification to Maxwell’s electrodynamics as an attempt to remove the divergence in the self-energy of the electron~\cite{BI1934}. The corresponding Born-Infeld (BI) Lagrangian is given by
\begin{equation}
    \mathcal{L} = b^2 \left(1-\sqrt{1 + \frac{1}{b^2}(\mathcal{F} - \mathcal{G}^2)} \right),
    \label{eq:LagrangeBI}
\end{equation}
where $b$ is a constant with dimensions of field strength, $\mathcal{F}\equiv F_{\mu\nu}F^{\mu\nu}/4$, and $\mathcal{G}\equiv \tilde{F}_{\mu\nu}F^{\mu\nu}/4$. Coupling BI electrodynamics to gravity gives rise to nontrivial BH solutions~\cite{Pellicer:1969cf, Salazar:1987ap, Breton2010}. Moreover, photon--photon scattering effects in quantum electrodynamics suggest that electrodynamics may become nonlinear at sufficiently high energies~\cite{Tommasini:2008lrs, Pike:2014wha, Gaete:2014nda, Gaete:2022lkf, Kadlecova:2023qxn}. 


Motivated by these considerations, Kruglov~\cite{2Kruglov2017, Kruglov2017} proposed a generalized form of BI electrodynamics that phenomenologically incorporates vacuum birefringence effects,
\begin{equation}
         \mathcal{L} = \frac{1}{\beta} \left[ 1 - \left(1 + \frac{\beta \mathcal{F}}{q} - \frac{\beta \gamma \mathcal{G}^2}{2q} \right)^{q}\right],
\label{LagrangeKruglovFull}
\end{equation}
where $\beta$ and $\gamma$ are constants with dimensions $[L]^4$, and $q\in\mathbb{R}$ is a dimensionless parameter. This construction interpolates continuously between Maxwell electrodynamics and BI-type nonlinear regimes while remaining formally well-defined over a broad parameter range\footnote{However, it has been shown that, for a similar form of NLED given in Eq.~\eqref{LagrangeKruglovFull}, causality considerations impose constraints on the allowed parameter space, $1/2 \le q < 1$~\cite{Russo:2024kto}. The implications of NLED causality for the BH structure have been discussed recently in Ref.~\cite{Russo:2026vnj}.}.


A crucial feature of NLED is that photon propagation is not governed by the background metric itself, but by an effective geometry determined by the electromagnetic field configuration. This was first demonstrated by Novello {\it et al.}~\cite{Novello2000}, who showed that photons in NLED follow null geodesics of an effective metric. Consequently, optical observables may substantially differ from those expected from the background geometry alone. In Ref.~\cite{AH2020}, it was shown that the effective geometry associated with Kruglov electrodynamics can admit stable photon orbits outside the event horizon, indicating the presence of qualitatively distinct null structures induced purely by nonlinear electromagnetic effects. More recently, the impact of effective photon geometries on black hole shadows and strong gravitational lensing has been investigated in several NLED-inspired spacetimes~\cite{HSR2023,Murk:2024nod,Guzman-Herrera:2024fkg,KumarWalia:2024yxn,Tlemissov:2025nnk}.


According to the results of Ref.~\cite{AH2020}, the purely magnetic sector of Eq.~\eqref{LagrangeKruglovFull} with small $\beta$ admits analytic BH solutions whose spacetime geometry closely approximates that of the Reissner--Nordström (RN) solution outside the event horizon, even near extremality and for arbitrary values of $q$. In contrast, the induced effective geometry governing photon propagation remains highly sensitive to the parameter $q$. As a result, while the motion of massive particles remains nearly indistinguishable from that in the standard RN spacetime, photon trajectories can undergo significant modifications. This behavior differs from several other NLED models, such as the ABG~\cite{Ayon-Beato:2000mjt}, Bronnikov~\cite{Bronnikov:2000vy,Bronnikov:2017sgg}, and ModMax theories~\cite{Flores-Alfonso:2020euz}, where deviations in the spacetime geometry itself significantly affect the motion of all particle species. The Kruglov model therefore provides a useful framework for isolating the effects of the effective photon geometry on null geodesics and associated optical observables while leaving the background spacetime geometry largely unchanged. 


In this work, we investigate the effective null geodesic structure associated with BHs in Kruglov nonlinear electrodynamics and its consequences. We focus particularly on the role of the parameter $q$ in shaping photon trajectories, photon spheres, and relativistic image formation. Using fully numerical calculations, we analyze light deflection, BH shadows, and accretion-disk images within the effective geometry framework. This paper is organized as follows. In Sec.~\ref{sec:Kruglov} we review Kruglov electrodynamics and the corresponding BH solution. In Sec.~\ref{sec. null geodesic} we discuss the effective geometry and the associated photon spheres. In Sec.~\ref{sec:deflection} we analyze photon deflection and null trajectories. In Sec.~\ref{sec:shadow} we study the BH shadow and accretion-disk images, and compare the resulting shadow radius with current horizon-scale observational constraints. Finally, we summarize our results in Sec.~\ref{sec:conc}.

\section{Kruglov Born-Infeld-type electrodynamics}
\label{sec:Kruglov}

Consider the purely-magnetic form of the Lagrangian~\eqref{LagrangeKruglovFull}, 
\begin{equation}
    \mathcal{L}_K = \frac{1}{\beta} \left[ 1 - \left(1 + \frac{\beta \mathcal{F}}{q} \right)^{q} \right],
    \label{LagrangeKruglovcok}
\end{equation}
which interpolates between several well-known NLED models. In particular, the choice $q=1/2$ reproduces the BI Lagrangian, while $q=1$ yields standard Maxwell electrodynamics. In the limit $q\to\infty$, the model approaches exponential electrodynamics as discussed by Hendi and Sheyki~\cite{Hendi:2013mka},
\begin{equation}
    \mathcal{L}\rightarrow\frac{1}{\beta} \left[ 1 - \exp\left(\beta\mathcal{F}\right)\right].
\end{equation}

 A convenient gauge choice consistent with static and spherically-symmetric configuration corresponds to the magnetic monopole potential
\begin{equation}
A_t=A_r=A_\theta=0,\qquad
A_\phi=Q(1-\cos\theta),
\end{equation}
where $Q$ is the magnetic charge. The only non-vanishing component of the electromagnetic tensor is
\begin{equation}
F_{\theta\phi}=Q\sin\theta,
\end{equation}
from which the Maxwell invariant becomes
\begin{equation}
\mathcal F=\frac14F_{\mu\nu}F^{\mu\nu}
=\frac12F_{\theta\phi}F^{\theta\phi}
=\frac{Q^2}{2r^4}.
\end{equation}
The corresponding energy-momentum tensor is
\begin{align}
T_{\mu\nu}&=2\mathcal{L}_{\mathcal{F}}F_{\mu\alpha}F_{\nu}^{\alpha}-g_{\mu\nu}\mathcal{L}\nonumber\\
&=\frac{g_{\mu\nu}}{\beta} \left[ 1 - \left(1 + \frac{\beta \mathcal{F}}{q} \right)^{q} \right]-2F_{\mu\alpha}F_{\nu}^{\alpha}\left(1+{\beta\mathcal{F}\over q}\right)^{q-1},
\end{align}
where $\mathcal{L}_{\mathcal{F}}\equiv\partial\mathcal{L}/\partial\mathcal{F}$.

For $q<0$, the quantity $1+\beta\mathcal{F}/q$ vanishes at the critical radius
\begin{equation}
r_z\equiv
\left(
\frac{\beta Q^2}{2|q|}
\right)^{1/4}.
\end{equation}
At this radius, the Lagrangian remains finite, $\mathcal L(r_z)=1/\beta$, whereas its derivatives become non-analytic. The NLED field ceases to be well-defined. For $r<r_z$ the fractional power acquires a complex value, rendering both the Lagrangian and the corresponding energy-momentum tensor complex. Thus $r=r_z$ acts as the natural boundary of the real branch of the Kruglov theory for negative values of $q$, and throughout this work we restrict our analysis to the real branch of the solution ($r>r_z$).

\begin{figure}[htbp]
	\centering
\includegraphics[width=0.7\textwidth]{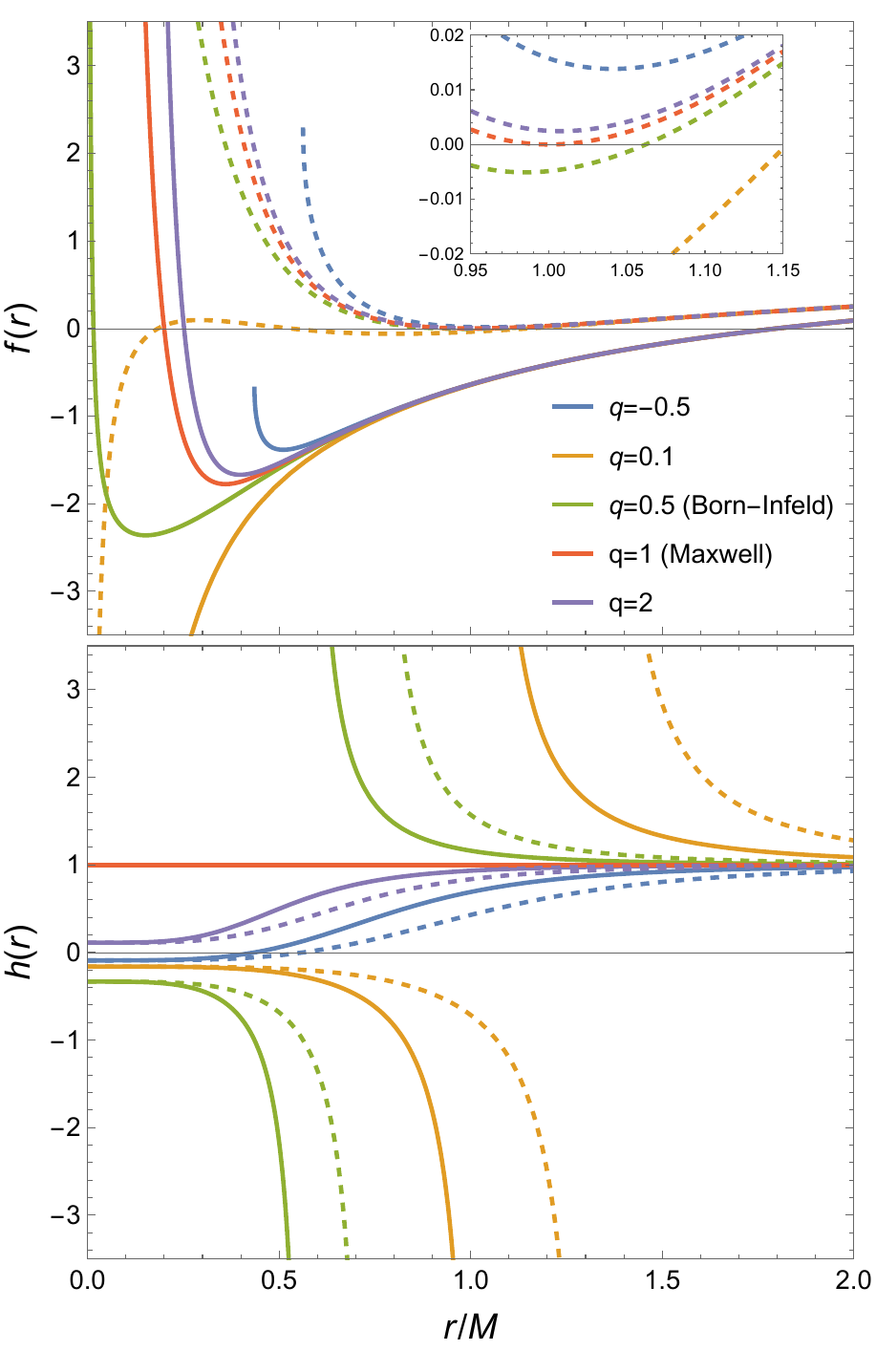}
	\centering
	\caption{(Top) The Kruglov metric function $f(r)$ and (Bottom) photon effective geometry correction $h(r)$ with $Q/M=0.6$ (solid lines) and $Q/M=1$ (dashed lines). We set $\beta/M^2=0.1$. Outside the event horizon, with the variation of $q$, the function $f(r)$ barely deviates, while the correction term $h(r)$ has slight modifications from the Maxwell electrodynamics. At $q=0.1$, $h(r)$ diverges outside the horizon, i.e $r_p>r_H^+$.}
 \label{kruglovplot}
\end{figure}

The corresponding BH geometry is given by the line element
\begin{equation}
	ds^2=-f(r)dt^2+\frac{1}{f(r)}dr^2+r^2d\Omega^2,
\label{line element ori}
\end{equation}
with metric function~\cite{2Kruglov2017, Kruglov2017},
\begin{align}
	f(r) &=1-{2\over r}\int T_0^0\,r^2\,dr\nonumber\\
	&= 1 - \frac{2M}{r}-\frac{2 r^2}{3 \beta} \left[_2F_1 \left(-\frac{3}{4}, -q; \frac{1}{4}; -\frac{Q^2 \beta}{2qr^4} \right) -1\right],
    \label{fr}
\end{align}
where $M$ is the BH mass and $_2F_1 \left(a,b;c;z \right)$ denotes the hypergeometric function of its arguments. For $q>0$, the argument of the hypergeometric function,
\[
z\equiv-\frac{\beta Q^2}{2qr^4},
\]
is always negative and the metric function remains real for all $r>0$. For $q<0$, however,
\[
z=\frac{\beta Q^2}{2|q|r^4},
\]
which exceeds unity when $r<r_z$. Since the Gaussian hypergeometric function possesses a branch cut along the real axis for $z\ge1$, the metric function generally acquires an imaginary part inside this region. 

Expanding the Gaussian hypergeometric function in Eq.~\eqref{fr} around $\beta=0$ to third order yields
\begin{equation}
    f(r) \approx 1-\frac{2 M}{r}+\frac{Q^2}{
   r^2}+\frac{\beta Q^4 (q-1)}{20 q r^6} +\frac{\beta ^2 Q^6 (q-2)
   (q-1)}{216 q^2 r^{10}} + \mathcal{O}(\beta^3)\label{expandf}
\end{equation}
which reduces to the standard RN solution when $q=1$. This asymptotic expansion provides an accurate approximation for sufficiently small values of $\beta$ outside the event horizon~\cite{2Kruglov2017,Kruglov2017}, or equivalently whenever
\[
\left|\frac{\beta Q^2}{2qr^4}\right|<1,
\]
which is typically satisfied outside the event horizon. It should be emphasized that Eq.~\eqref{expandf} represents only the real analytic branch of the exact solution\footnote{In particular, for $q<0$ the exact metric function~\eqref{fr} develops a branch cut when $r<r_z$, causing the Gaussian hypergeometric function to acquire an imaginary part. This non-analytic behavior cannot be captured by the Taylor expansion, which is intrinsically restricted to the region $|z|<1$.}. The behavior of the exact metric is illustrated in Fig.~\ref{kruglovplot} and discussed in detail in Ref.~\cite{AH2020}.

A crucial feature of NLED is that photons do not propagate along the null geodesics of the background spacetime. Instead, within the geometric-optics approximation, the propagation of electromagnetic discontinuities is governed by the following effective geometry~\cite{Novello2000},
\begin{equation}
g_{\rm eff}^{\mu\nu}=\mathcal L_{\mathcal F}g^{\mu\nu}-4\mathcal L_{\mathcal FF}F^\mu{}_{\alpha}F^{\alpha\nu}.
\label{eq:novello}
\end{equation}
For the Kruglov Lagrangian~\eqref{LagrangeKruglovcok}, one obtains
\begin{align}
\mathcal L_{\mathcal F}&=-\left(1+{\beta\mathcal{F}\over q}\right)^{q-1}=-\left(1+\frac{\beta Q^2}{2qr^4}\right)^{q-1},\nonumber\\ 
\mathcal L_{\mathcal FF}&={\beta\left(1-q\right)\over q}\left(1+{\beta\mathcal{F}\over q}\right)^{q-2}=-\frac{\beta(q-1)}{q}\left(1+\frac{\beta Q^2}{2qr^4}\right)^{q-2}.
\end{align}

For the purely magnetic monopole configuration, the nonlinear correction contributes anisotropically to the effective geometry. The temporal and radial components differ from those of the background spacetime only through the common conformal factor $\mathcal L_{\mathcal F}$. The nontrivial modification therefore arises in the angular sector. Using
\begin{equation}
F^\theta{}_\alpha F^{\alpha\theta}=F^\phi{}_\alpha F^{\alpha\phi}=-\frac{Q^2}{r^6}=-{2\mathcal{F}\over r^2},
\end{equation}
the angular component of the effective metric becomes
\begin{align}
g_{\rm eff}^{\theta\theta}&=-\frac1{r^2}\left(\mathcal L_{\mathcal F}-8\mathcal F\,\mathcal L_{\mathcal FF}\right)=-\frac1{r^2}\left(1+\frac{\beta Q^2}{2qr^4}\right)^{q-2}\left[1+\frac{(8q-7)\beta Q^2}{2qr^4}\right],\nonumber\\
g_{\rm eff}^{\varphi\varphi}&=g_{\rm eff}^{\theta\theta}\sin^2\theta.
\label{eq:gtheta}
\end{align}
The angular geometry is governed by the combination
\(
\mathcal L_{\mathcal F}-8\mathcal F\mathcal L_{\mathcal FF}.
\)

Removing the overall conformal factor, the effective line element can be written as
\begin{equation}
ds_{\rm eff}^2=-f(r)\,dt^2+\frac{dr^2}{f(r)}+h(r)r^2d\Omega^2,
\label{effline}
\end{equation}
where
\begin{equation}
h(r)\equiv\frac{2qr^4+\beta Q^2}{2qr^4+\beta(8q-7)Q^2}.
\label{hr}
\end{equation}
Eq.~\eqref{effline} shows that all nontrivial modifications to photon propagation are encoded in the angular correction factor $h(r)$. In particular, the denominator of $h(r)$ is proportional to the quantity
\(
\mathcal L_{\mathcal F}
-
8\mathcal F\mathcal L_{\mathcal FF},
\)
such that the effective optical geometry becomes degenerate whenever
\begin{equation}
\mathcal L_{\mathcal F}-8\mathcal F\mathcal L_{\mathcal FF}=2qr^4+\beta(8q-7)Q^2=0.
\end{equation}
Since $\beta,\,Q^2>0$, and a physical radius requires $r_p^4\geq0$, this condition is satisfied only for
\begin{equation}
0<q<7/8,
\end{equation}
giving rise to the critical radius
\begin{equation}
r_p\equiv\left[\frac{\beta (7-8q)Q^2}{2q}\right]^{1/4},
\label{eq. rp pole}
\end{equation}
which therefore defines the parameter range for which the effective optical geometry develops a pole.

This behavior is illustrated in the lower panel of Fig.~\ref{kruglovplot}, where $h(r)$ diverges at $r=r_p$ for $0<q<7/8$. The function $h(r)$ also has zeros for negative $q$ located at $r=r_z$, which, in fact, coincides with the location where $f(r)$ become complex. An additional limiting case arises as $q\to0^+$, where the two-sphere element becomes negative and is effectively rescaled by a factor of $1/7$, indicating a substantial modification of the effective photon geometry. However, taking the limit $\lim_{q\to0^+}\mathcal{L}_K$ in Eq.~\eqref{LagrangeKruglovcok} reduces the NLED Lagrangian to zero, corresponding to the absence of an electromagnetic field.

It is important to emphasize that this pole is not a curvature singularity of the underlying spacetime, since the metric function $f(r)$ remains finite at $r=r_p$. Rather, it corresponds to a degeneracy of the optical geometry governing photon propagation. At this radius, the angular coefficient of the effective metric vanishes, causing the geometric-optics description to become singular. The effective metric should be interpreted as an approximate optical geometry whose applicability may break down in the immediate vicinity of $r=r_p$, even though the underlying spacetime itself remains regular.

\section{Photon geodesics}
\label{sec. null geodesic}

The motion of a test particle is governed by the geodesic equation,
\begin{equation}
    \frac{d\dot{x}^{\mu}}{d\tau} + \Gamma^{\mu}_{\alpha\beta} \dot{x}^{\alpha} \dot{x}^{\beta} = 0,
    \label{eq. geodesic equation}
\end{equation}
where $\tau$ is an affine parameter, $x^{\mu}=(t,r,\theta,\phi)$ and $\dot{x}^{\mu}\equiv dx^{\mu}/d\tau$. 
Restricting motion to the equatorial plane ($\theta = \pi/2,\, \dot{\theta} = 0$), one obtains two conserved quantities,
\begin{equation}
    f(r)\dot{t} = E, \qquad h(r)r^2\dot{\phi} = L,
    \label{eq. const of motion}
\end{equation}
which can be identified as the test particle’s conserved energy $E$ and angular momentum $L$. The four-velocity normalization condition is given by
\begin{equation}
    -f(r)\dot{t}^2 + \frac{1}{f(r)}\dot{r}^2 + h(r)r^2\dot{\phi}^2 = \epsilon,
    \label{eq. 4-velocity condition}
\end{equation}
where $\epsilon=0$ for photons and $\epsilon=-1$ for massive particles. 

Focusing on photon trajectories ($\epsilon=0$), substituting the constants of motion in Eq.~\eqref{eq. const of motion} into Eq.~\eqref{eq. 4-velocity condition} yields
\begin{equation}
\left(\frac{dr}{d\tau}\right)^2+V(r)L^2=E^2,\qquad V(r) \equiv \frac{f(r)}{h(r)r^2},
\label{eq. geodesic T+V=E}
\end{equation}
where $V(r)$ is known as the photon effective potential~\cite{Claudel:2000yi,Perlick:2004tq}. Applying the chain rule to Eq.~\eqref{eq. geodesic T+V=E} implies
\begin{equation}
    \frac{dr}{d\phi} = h(r)r^2 \sqrt{{1\over b^2} - V(r)},
    \label{trajectory1}
\end{equation}
where $b\equiv L/E$ is the impact parameter. Eq.~\eqref{eq. geodesic T+V=E} shows that the modification induced by nonlinear electrodynamics enters the photon dynamics entirely through the effective potential. In the Maxwell limit, $h(r)=1$. Any photon trajectory deviation rom the Maxwell case can be traced to the behavior of the effective optical geometry through $h(r)$.

Alternatively, the orbit equation may be written as
\begin{equation}
    \frac{d\phi}{dr} = \frac{1}{h(r)r^2} \sqrt{\frac{1}{1/b^2 - V(r)}},
    \label{eq. dphidr}
\end{equation}

\begin{figure}[htbp]
	\centering
\includegraphics[width=0.7\textwidth]{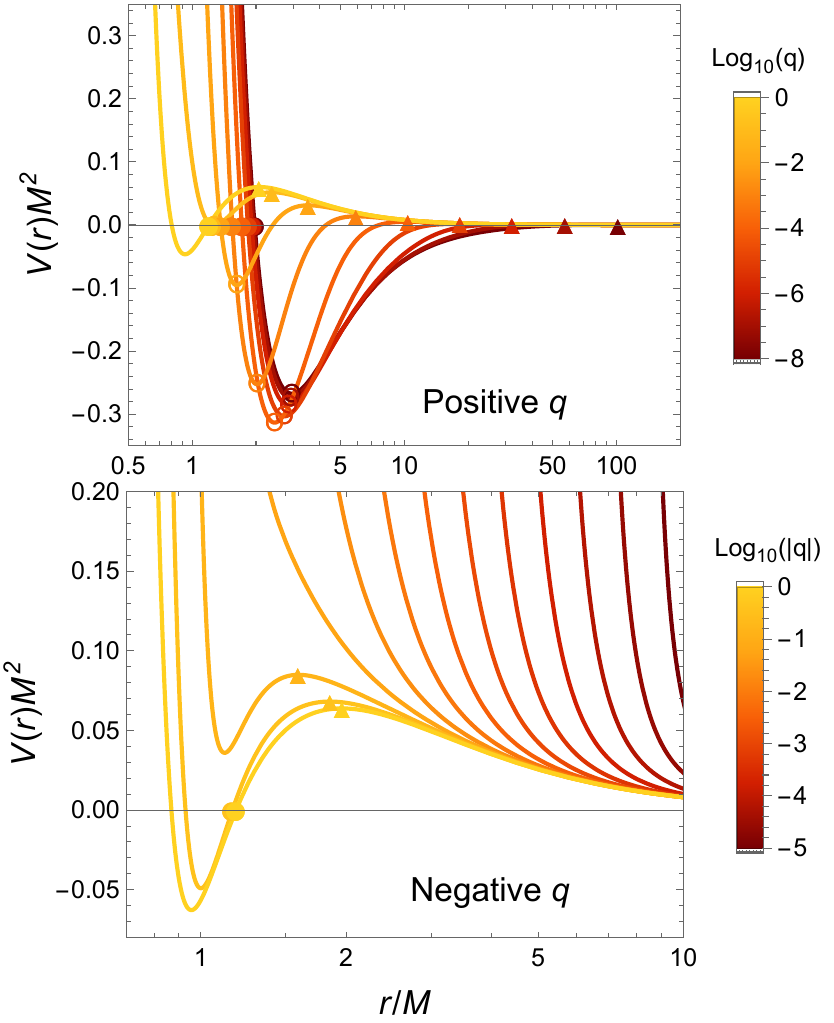}
	\centering
	\caption{Photon effective potential in Kruglov spacetime for (top) positive and (bottom) negative values of $q$, with $Q/M=0.98$. Solid circles, open circles, and triangles denote the locations of the outer horizon $r_H^+$, the stable photon sphere $r_{ps}^-$, and the unstable photon sphere $r_{ps}^+$, respectively. For certain values of $q>0$, a stable photon sphere appears between the horizon and the unstable one. For $q<0$, the effective potential diverges at $r_p$, and for some parameter ranges the spacetime has no event horizons.}
 \label{Veff}
\end{figure}

\begin{figure*}[htbp]
	\centering
\includegraphics[width=1\textwidth]{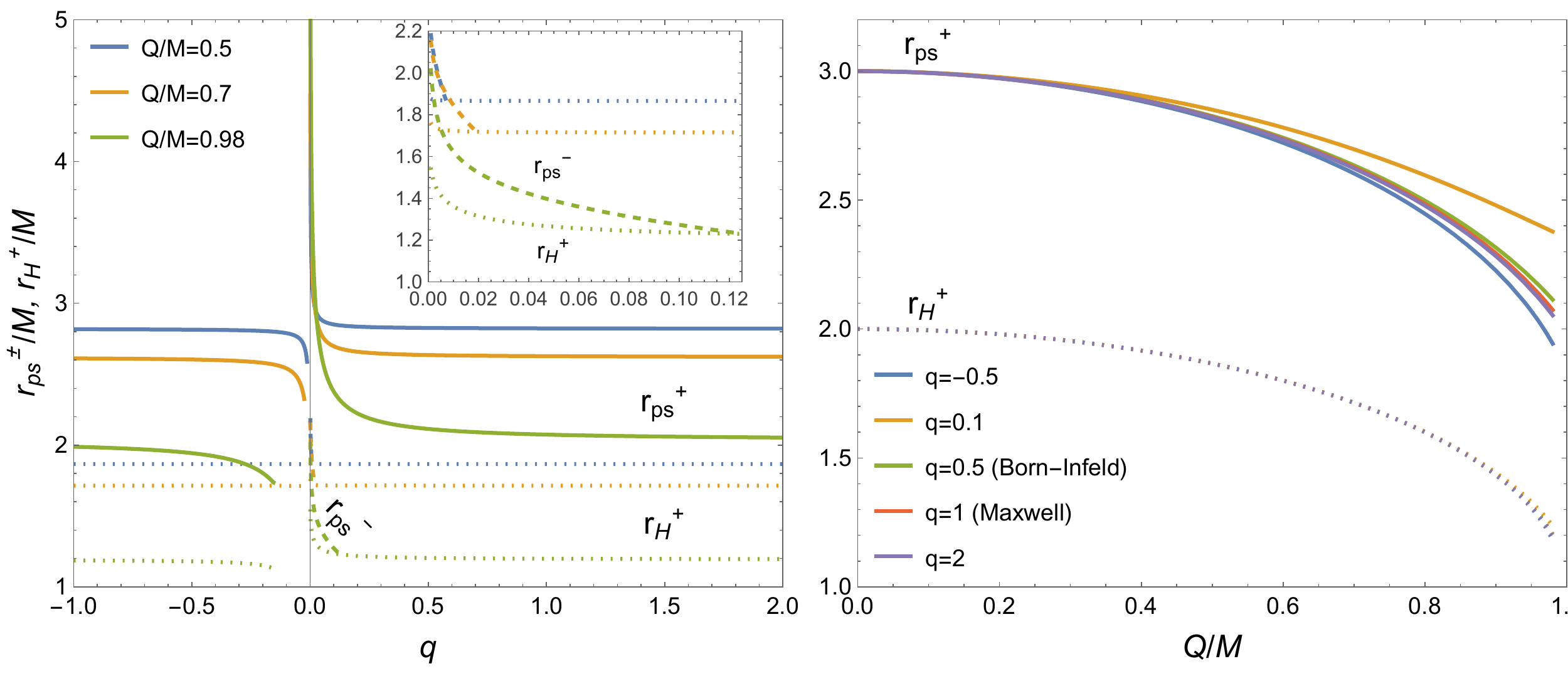}
	\centering
	\caption{Radius of the unstable photon sphere $r_{ps}^+$ (solid lines), stable photon sphere $r_{ps}^-$ (dotted lines), and the outer horizon $r_H^+$ (dotted lines) shown as a function of $q$ (left) and $Q$ (right). In the left figure, stable photon spheres are shown only when $r_H^+<r_{ps}^-<r_{ps}^+$. The curve for negative $q$ is also cut off at the absence of horizon. In the limit $q\to0^+$, $r_{ps}^-$ asymptotically reaches $\sim3M$.}
 \label{rps}
\end{figure*}

The photon sphere corresponds to a circular null orbit at radius $r_{ps}$, obtained by imposing the conditions
\begin{equation}
\frac{dr}{d\phi}\bigg|_{r=r_{ps}}=
\frac{d^2r}{d\phi^2}\bigg|_{r=r_{ps}} = 0,
\end{equation}
which is equivalent to extremizing the effective potential
\begin{equation}
    \partial_r V(r_{ps}^\pm) = 0, \label{eq. rps}
\end{equation} 
with stability determined by $\partial_{rr}V(r_{ps}^+)>0$ or $\partial_{rr}V(r_{ps}^-)<0$. Due to the pressence of a special function on the metric, the photon sphere radius is computed by solving Eq.~\eqref{eq. rps} numerically using a generic numerical root finding method in our following discussion. The outer photon sphere $r_{ps}^+$ is typically unstable, while the inner photon sphere $r_{ps}^-$, when it exists, generally lies inside or near the horizon and is not observable~\cite{Cardoso:2008bp}. However, in some parameter space, the Kruglov BH exhibits a complicated structure.


Before discussing the numerical results, it is instructive to understand qualitatively how the structure of the effective potential is controlled by the correction factor $h(r)$. Depending on the value of the parameter $q$, three qualitatively different situations may arise:
\begin{enumerate}[label=(\roman*)]
\item $h(r)$ remains regular everywhere outside the event horizon,
\item $h(r)$ develops a pole,
\item $h(r)$ possesses a zero.
\end{enumerate} 
Since the effective potential is inversely proportional to $h(r)$, these three cases lead to distinct classes of photon trajectories and photon-sphere structures.

For finite values of $q$, both the background metric and the correction factor satisfy
\[
f(r)\rightarrow1,\qquad
h(r)\rightarrow1,
\]
as $r\rightarrow\infty$, so that the effective potential retains the qualitative structure of the RN spacetime at large distance. Only quantitative shifts in the position and height of the potential barrier occur, resulting in moderate changes in the unstable photon sphere.

The corresponding effective potentials are shown in Fig.~\ref{Veff}. For sufficiently small positive values of $q$, Eq.~\eqref{eq. rp pole} predicts that the correction factor develops a pole outside the event horizon. Since
\[
V(r)\propto\frac{1}{h(r)},
\]
the effective potential is driven to zero at the pole. Physically, this corresponds to a strong suppression of the effective centrifugal barrier experienced by photons. Consequently, the potential develops a local maximum outside the pole together with a local minimum between the pole and the event horizon. The former corresponds to the usual unstable photon sphere, whereas the latter gives rise to a stable circular photon orbit. As illustrated in the upper panel of Fig.~\ref{Veff}, this additional minimum becomes increasingly pronounced as $q$ decreases.

As $q\rightarrow0^+$, the correction factor approaches
\[
h(r)\rightarrow-\frac17,
\]
so that 
\[
V(r)\rightarrow
-\frac{7f(r)}{r^2}.
\]
The effective potential therefore becomes the reflected counterpart of the Schwarzschild potential with respect to the horizontal axis. Consequently, the unstable Schwarzschild photon sphere at $r=3M$ is transformed into a stable orbit in the effective optical geometry, while the unstable branch moves to arbitrarily large radii.

A different behavior occurs for negative values of $q$. In this case, the correction factor possesses a zero rather than a pole, causing the effective potential to diverge. Consequently, photon turning points always satisfy
\[
r>r_z,
\]
preventing photon trajectories from entering the region where the background metric becomes complex. The effective optical geometry therefore naturally excludes the complex domain from photon propagation. For massive particles, however, no analogous divergence exists in the corresponding effective potential, so additional analysis would be required to determine whether their trajectories can approach the complex region.




The existence of pole and the sign change of $h(r)$ outside the event horizon also leads to an interesting phenomenon for the photon geodesics. Since the photon effective potential is inversely proportional to $h(r)$, at the pole $r=r_{z}$ one has $d\phi/d\tau=d\phi/dr=0$. Below $r_z$, the $h(r)$ becomes negative, and hence the photon angular direction reverses. It corresponds to the photon turning point. If the pole is outside the event horizon, then there will be observational consequences, especially if the observable is probed beyond the pole location.

The qualitative deformation of the effective potential discussed above is directly reflected in the location of the circular photon orbits. Fig.~\ref{rps} summarizes the evolution of the stable and unstable photon-sphere radii as functions of $q$ and $Q$. In the left panel of Fig.~\ref{rps}, we show the branching of the photon spheres for varying $q$ in the near-extremal configuration with $Q/M=0.98$, focusing on photon spheres located outside the BH horizon. It can be seen that for sufficiently small $q$, there exists a stable photon sphere at $r=r_{ps}^-$. Both the stable and unstable photon sphere radii become larger as $q$ decreases. In the limit $q\to0^+$, the unstable photon sphere diverges to large distances, whereas the stable photon sphere converges to $3M$. This is due to the fact that the effective potential “flips” upside down compared to the Schwarzschild counterpart. On the other hand, negative values of $q$ show the opposite behavior, where smaller (more negative) $q$ results in a smaller photon sphere radius. There is also no stable photon sphere appearing outside the BH horizon for negative $q$, as expected from the previous analysis of the effective potential. Furthermore, the photon sphere and horizon radii converge in the limit of large absolute values of $q$. We also show the photon sphere behavior as a function of the charge $Q$ in the right panel of Fig.~\ref{rps}. We find that the photon sphere radius is more \textit{insensitive} to changes in $Q$ for smaller positive values of $q$. On the other hand, small negative values of $q$ result in the opposite behavior.


The critical impact parameter $b_c$, corresponding to the unstable photon orbit, is obtained by evaluating Eq.~\eqref{trajectory1} at $r=r_{ps}^+$,
\begin{equation}
    b_c=\sqrt{\frac{1}{V(r_{ps}^+)}}.
    \label{eq. crit impact parameter bc}
\end{equation}
Photons with $b=b_c$ asymptotically spiral towards the photon sphere, those with $b>b_c$ scatter back to infinity, and those with $b<b_c$ are absorbed by the BH horizon. Thus, the photon sphere and its associated impact parameter determine the observable shadow boundary of the BH~\cite{Synge:1966okc, Bardeen:1973tla, Hod:2013jhd}. 

It should be emphasized that the shadow boundary is determined exclusively by the unstable photon sphere through the critical impact parameter $b_c$. The stable photon sphere does not contribute directly to the shadow edge. Instead, its observational signature arises from photons executing multiple revolutions before escaping towards the observer. Consequently, the presence of a stable photon orbit is expected to affect higher-order lensed images and the photon-ring structure rather than the shadow diameter itself.




It is important to distinguish the notion of stability discussed in the present work from the dynamical stability of the spacetime itself. Here, stability refers exclusively to the existence of local minima of the effective potential governing null geodesics in the effective optical geometry. A complete assessment of the physical stability of these configurations would require analyses beyond the geometric-optics approximation, including perturbations of the nonlinear electromagnetic field, hyperbolicity of the field equations, and the causal structure of the underlying theory. Such investigations lie beyond the scope of the present work.

\section{Light deflection}
\label{sec:deflection}

\begin{figure*}[htbp]
	\centering
	\includegraphics[width=1\textwidth]{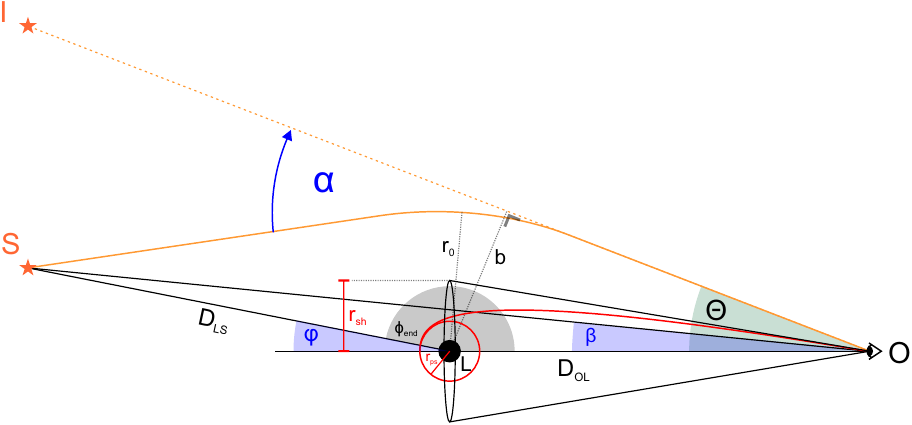}
	\centering
	\caption{Schematic diagram of light deflection by a BH acting as a lens ($L$). Points $S$ and $O$ are the locations of the source and observer, respectively, while $I$ is the apparent position of the source as seen by the observer. The distances between the source, lens, and observer are given by $D_{LS}$ and $D_{OL}$, respectively. The quantity $r_0$ denotes the distance of closest approach of the light to the lens, and is related to the impact parameter $b$ through the condition $V(r_0)=1/b^2$. The deflection angle is denoted by $\alpha$. The red curves correspond to photon trajectory with the critical impact parameter, asymptotically orbiting the BH at the photon sphere radius $r_{ps}$.}
        \label{figdef}
\end{figure*}

The geometry of gravitational lensing is shown in Fig.~\ref{figdef}. A source at $S$, located at a distance $D_{LS}$ from the lens $L$ (in this case, a BH), sends a light ray toward the lens with an impact parameter $b$. The lens deflects the light ray by an angle $\alpha$, and the ray travels toward the observer $O$, located at a distance $D_{OL}$ from the lens. As a consequence, from the observer’s point of view, the source appears to be located at $I$ instead of $S$.

Now, we would like to solve for the deflection angle $\alpha$. There is a particular approximation method originally proposed by Bozza for computing the light deflection angle in the Schwarzschild geometry in the strong-field limit~\cite{Bozza2002}. The method was then generalized by Tsukamoto in Ref.~\cite{Tsukamoto2017} for arbitrary spherically symmetric and asymptotically flat spacetimes. These methods are particularly useful for the direct calculation of the deflection angle for a given impact parameter, since one can compute the lensing coefficients to obtain an approximate deflection angle for any impact parameter close to its critical limit.

Although this approach has been widely used in the literature (see, e.g., Refs.~\cite{Pereira:2025fvg,Cheong:2025lwp,Rodriguez:2025gfw,Kumar:2025mpb,Khodadi:2025upl} for some recent works), the approximation is too restrictive for our purposes; it only accurately approximates the deflection angle for impact parameters near the critical value $b_c$. Therefore, we instead consider a full numerical integration approach to compute the deflection angle $\alpha$. In addition, we present a comparison between the deflection angles obtained from full numerical integration and those from the strong-field approximation in Appendix~\ref{sec. strong field approx}.

Following a ray-tracing procedure, we integrate the differential equation given by Eq.~\eqref{eq. dphidr}. The integration begins from the observer at $D_{OL}$ toward the BH, up to the point of closest approach $r_0$, where $V(r_0)=1/b^2$. From there, the integration is continued to the source located at $D_{LS}$. It should be noted that this integration is only valid when $b\geq b_c$. This integration gives the quantity $\phi_{end}=\pi-\varphi$, and it can be written as\footnote{Due to the divergence of $d\phi/dr$ at $r_0$, the integration must be carried out carefully. Fortunately, it is sufficient to set $r_0$ at the integration boundary to $r_0+\epsilon$, where $\epsilon\ll1$. In our case, the integration values converge for smaller values of $\epsilon$, particularly for $b\gtrsim b_c$.}
\begin{equation}
    \phi_{end}=-\int_{D_{OL}}^{r_0}\left(\frac{d\phi}{dr}\right)dr+\int_{r_0}^{D_{LS}}\left(\frac{d\phi}{dr}\right)dr,\quad (b\geq b_c)
    \label{eq. phi end}
\end{equation}
where $d\phi/dr$ is given in Eq.~\eqref{eq. dphidr}. The minus sign in the first integral accounts for the fact that the integration is carried out backward. For photon trajectories with $b<b_c$, they are absorbed by the BH horizon, and the definition of $r_0$ by $V(r_0)=1/b^2$ is ill-defined. In this case, the closest distance of approach is the BH horizon itself; hence, $r_0=r_H$, and the integration reads
\begin{equation}
    \phi_{end}=-\int_{D_{OL}}^{r_H}\left(\frac{d\phi}{dr}\right)dr. \qquad (b< b_c)
    \label{eq. phi end horizon}
\end{equation}

According to Fig.~\ref{figdef}, for $b>b_c$, we know that
\begin{equation}
    \alpha=\Theta-\varphi+\gamma,
    \label{eq. alpha theta phi gamma}
\end{equation}
where $\Theta$ and $\gamma$ can be obtained by elementary geometry, yielding 
\begin{equation}
    \tan\Theta=\frac{b}{\sqrt{D_{OL}^2-b^2}},\qquad \sin\gamma=\frac{b}{D_{LS}}.
\end{equation}
Therefore, Eq.~\eqref{eq. alpha theta phi gamma} now reads
\begin{equation}
    \alpha=\arctan\left(\frac{b}{\sqrt{D_{OL}^2-b^2}}\right)+\arcsin\left(\frac{b}{D_{LS}}\right)+\phi_{end}-\pi,
    \label{eq. alpha tan sin}
\end{equation}
In the limit of a distant observer and a distant source, \textit{i.e.}, $D_{LS}\to\infty$ and $D_{OL}\to\infty$, the first two terms on the right-hand side of Eq.~\eqref{eq. alpha tan sin} vanish, and the generic expression of the deflection angle (see, e.g., Ref.~\cite{Weinberg:1972kfs})
\begin{equation}
    \alpha=2\int_{r_0}^{\infty}\left(\frac{d\phi}{dr}\right)dr -\pi
\end{equation}
is retrieved.

\begin{figure}[htbp]
	\centering
	\includegraphics[width=0.49
    \textwidth]{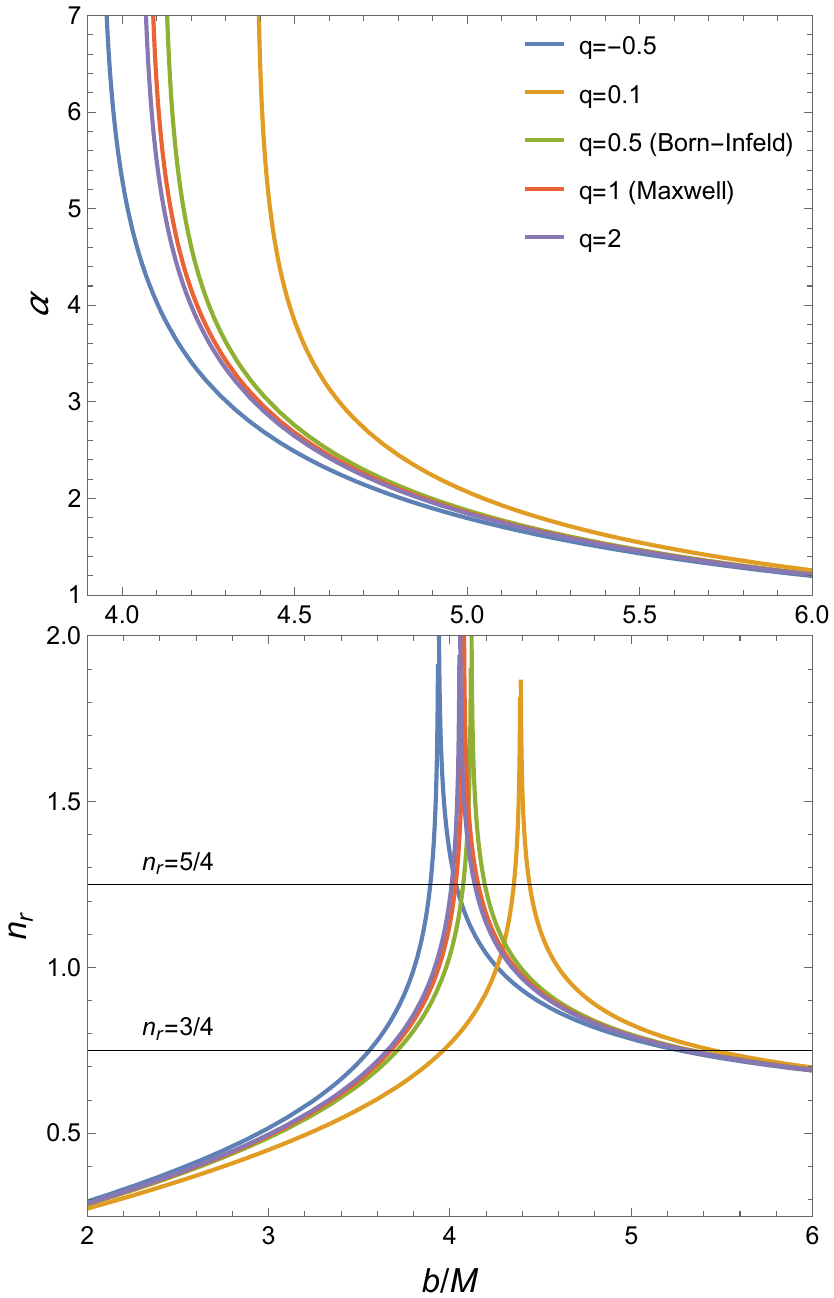}
	\centering
	\caption{(Top) Deflection angle $\alpha$ and (Bottom) number of photon orbits $n_r$ as functions of the impact parameter $b$ for a near-extremal BH with $Q/M=0.98$. The value of $q$ generally shifts the critical impact parameter at which both $\alpha$ and $n_r$ diverge. The numerical values of the impact parameter range and width for both lensed and photon ring trajectories can be found in Tab.~\ref{tab. b range}.}
        \label{figalpha}
\end{figure}

The resulting deflection angle as a function of the impact parameter for various values of $q$ is shown in the left panel of Fig.~\ref{figalpha}. In general, smaller positive values of $q$ result in a greater deflection angle, while negative values of $q$ exhibit the opposite behavior. These deviations are most significant near the critical impact parameter, where the deflection angle for all values of $q$ asymptotically converges to the same value at large deflection angles. This behavior is expected, since the deviations in the spacetime geometry (cf. Fig.~\ref{kruglovplot}) from Maxwell electrodynamics mainly occur near the BH horizon and approach the same geometry at large distances. Therefore, probing the values of $q$ through observations of photon geodesics must be performed in the strong-field regime.

We also examine the photon trajectories around the BH. They are classified into three categories (see, e.g., Refs.~\cite{Gralla:2019xty, Fauzi:2024nta,Zeng:2023fqy,Meng:2023htc,Macedo:2024qky}); The direct emission ($n_r \leq 3/4$), lensed emission ($3/4 < n_r \leq 5/4$), and light ring ($n_r > 5/4$), with $n_r$ is the number of orbits described as
\begin{equation}
    n_r = \frac{\phi_{end}}{2\pi},
    \label{eq. number orbits}
\end{equation}
with $\phi_{end}$ given by either Eq.~\eqref{eq. phi end} or Eq.~\eqref{eq. phi end horizon}, depending on the impact parameter.
\begin{figure*}[htbp]
	\centering
\includegraphics[width=0.9\textwidth]{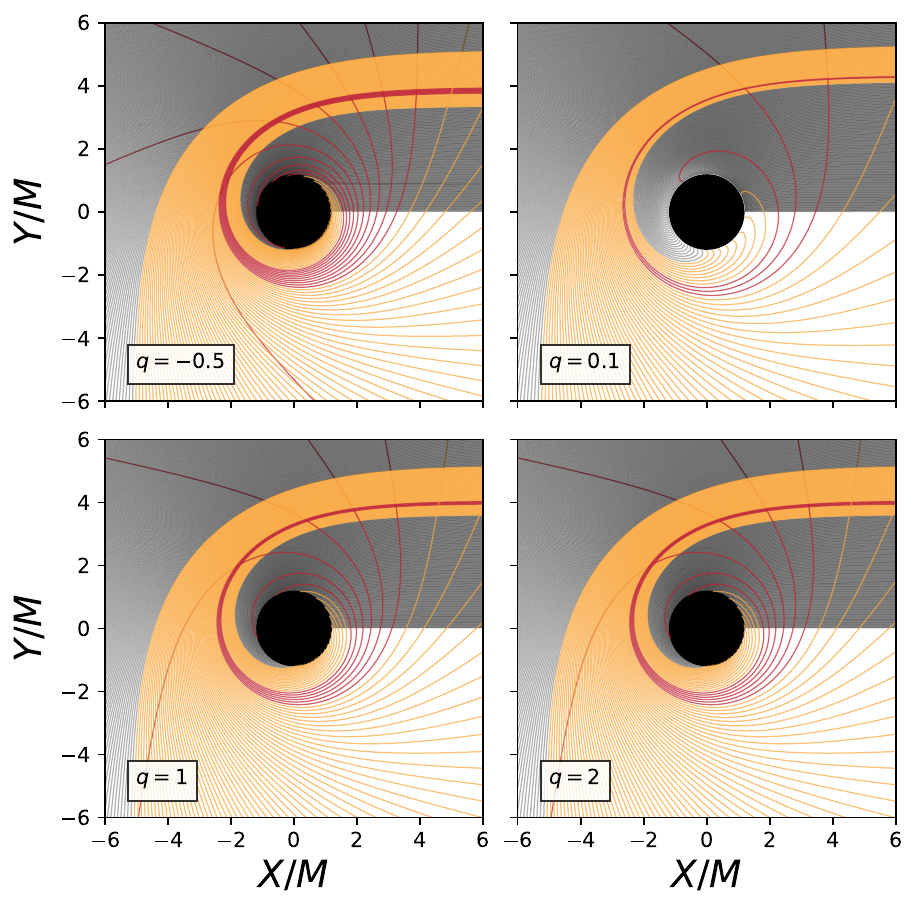}
	\centering
	\caption{Photon trajectories around a Kruglov BH for $Q/M=0.98$ with several values of $q$, where the black solid circle denotes the location of the outer horizon $r_H^+$. The gray, orange, and red curves represent light paths with direct, lensed, and light ring trajectories, respectively. At $q=0.1$, one may observe that the photon changes its angular direction near the BH horizon, which resembles (but is not) a frame dragging effect seen in rotating BH spacetimes.}
 \label{Shadowtrace}
\end{figure*}

The right plot in Fig.~\ref{figalpha} shows that, aside from the shift in the divergence of $n_r$ at the critical impact parameter, smaller positive values of $q$ correspond to a narrower range of impact parameters that give rise to light ring trajectories. In contrast, negative values of $q$ produce the largest deviation from Maxwell’s electrodynamics, yielding the widest range of light-ring impact parameters. We present the numerical values of the impact parameter ranges and their widths for both light ring and lensed emission trajectories in Tab.~\ref{tab. b range}.

\begin{table}[htbp!]
\centering
\setlength{\tabcolsep}{7pt}
\begin{tabular}{ccccc}
\hline\hline
\multirow{2}{*}{q} & \multicolumn{2}{c}{$n_r\geq3/4$} & \multicolumn{2}{c}{$n_r\geq5/4$} \\ \cline{2-5} 
& $b/M$ & $\Delta b/M$ & $b/M$ & $\Delta b/M$ \\ \hline
-0.5 & $3.56-5.26$ & $1.70$ & $3.89-4.03$ & $0.14$\\
0.1 & $3.96-5.45$ & $1.49$ & $4.35-4.44$& $0.09$\\
0.5 & $3.71-5.33$ & $1.62$ & $4.07-4.19$ & $0.12$ \\
1 & $3.67-5.31$ & $1.64$ & $4.03-4.16$ & $0.13$ \\
2 & $3.66-5.31$ & $1.65$ & $4.01-4.14$ & $0.13$\\
\hline
\end{tabular}
\caption{Numerical values of the impact parameter range $b$ and its width $\Delta b = b_{\max} - b_{\min}$ for both lensed ($n_{r}\geq 3/4$) and light ring ($n_{r}\geq 5/4$) trajectories for the values of $q$ presented in Fig.~\ref{figalpha} with $Q/M=0.98$.}
\label{tab. b range}
\end{table}

The analysis of the number of photon orbits plays a crucial role in understanding the image features of the BH. The range of impact parameters associated with light ring trajectories affects the observability of critical photon rings in BH images. A wider range of light ring impact parameters increases the set of photon trajectories that can undergo several orbits near the BH photon sphere before reaching a distant observer. Consequently, the critical photon ring becomes easier to observe. This implies that smaller positive values of $q$ make the photon ring feature in the BH image harder to detect for a distant observer, whereas negative values of $q$ have the opposite effect.

The photon trajectories with various impact parameters around the BH are shown in Fig.~\ref{Shadowtrace}. It can be seen that the width of the light ring impact parameter shown in Fig.~\ref{figalpha} is related to the width of the light ring trajectories shown by the red curves, where smaller values of $q$ lead to narrower light-ring trajectories, while negative values of $q$ exhibit the opposite behavior. The effect of negative $h(r)$ outside the horizon is clearly reflected in the photon trajectories with $q=0.1$: the photons change their angular direction near the BH exactly at the pole of $h(r)$. As discussed previously, this is a consequence of the divergence and sign change of $h(r)$ at $r=r_p$, causing the geodesic equation in Eq.~\eqref{eq. dphidr} to change sign and reverse the angular direction of the photon orbit. We will see shortly in the next section that this has consequences for BH imaging when the light source (or the accretion disk) extends up to the pole location at $r=r_p$.

\section{Shadow and accretion disk images}
\label{sec. shadow image}

A direct observable as a consequence of light deflection in the strong-field regime of BH spacetimes is the shadow of the BH horizon. Since BHs do not emit light directly, their observational signatures must be inferred from the propagation of radiation emitted by the surrounding matter. A standard and widely used choice of light source is an accretion disk. In this section, we compute both the BH shadow radius and the images of various thin accretion disks around the BH, with particular emphasis on examining the impact of the parameter $q$.

\subsection{Shadow radius}
\label{sec:shadow}

\begin{figure*}[htbp]
	\centering
	\includegraphics[width=1\textwidth]{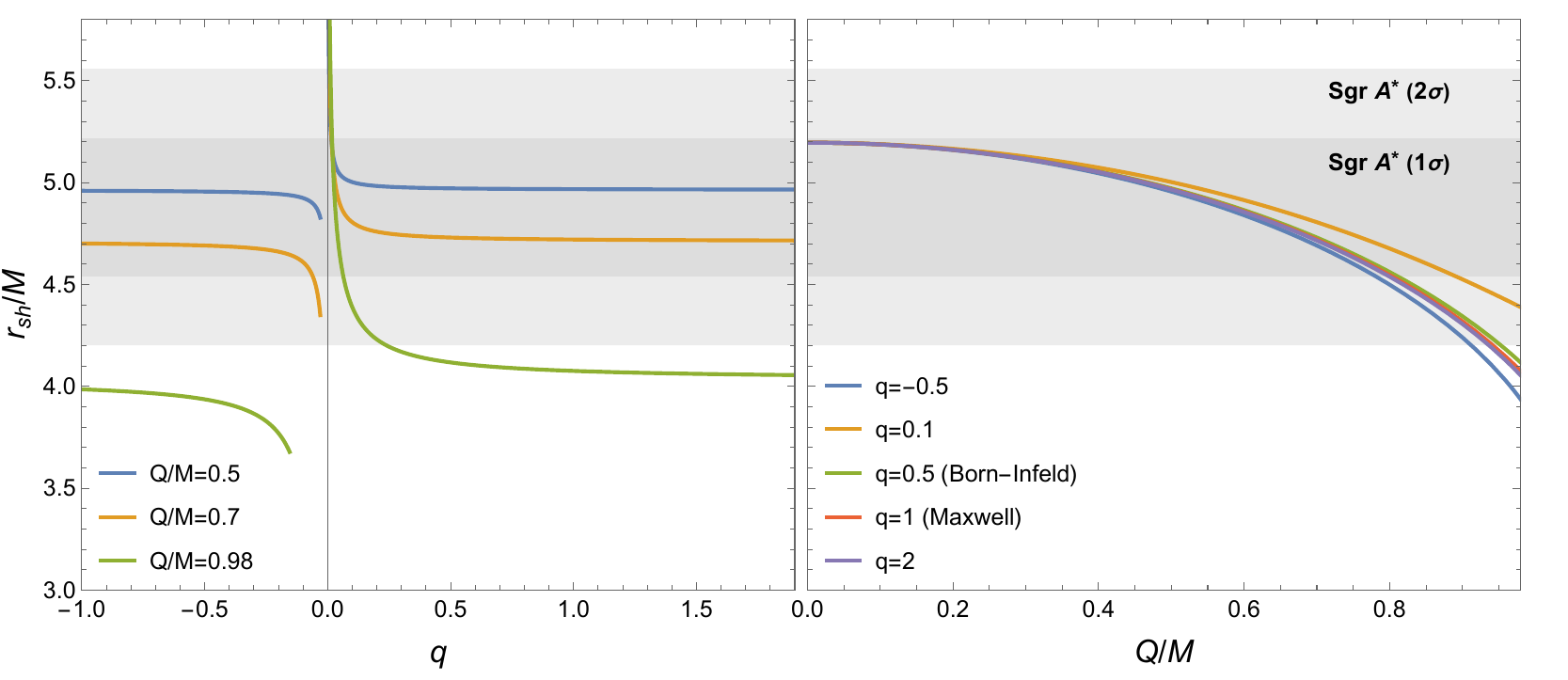}
	\centering
	\caption{The BH shadow radius as a function of $q$ (left) and $Q$ (right). The $1\sigma$ and $2\sigma$ constraints from Sgr A* observations are shown by the dark gray and light gray areas, respectively. The curve for negative $q$ is cut off in the absence of a horizon.}
        \label{rshad}
\end{figure*}

To calculate the shadow radius, we first consider the geometrical setup illustrated in Fig.~\ref{figdef}. From elementary geometry, the angular shadow angle, \textit{i.e.}, $\Theta$ when $b\to b_c$, is given by
\begin{equation}
	\tan \Theta|_{b\to b_c} = \frac{r_{sh}}{D_{OL}}= \left.\sqrt{\frac{h(r)r^2}{B(r)}}\frac{d\phi}{dr}\right|_{r={D_{OL}}}.
	\label{eq. tan theta}
\end{equation}
Substituting Eq.~\eqref{eq. dphidr} into Eq.~\eqref{eq. tan theta} with $b=b_c$, the explicit form of $r_{sh}$ reads
\begin{equation}
    r_{sh}=D_{OL}b_c\sqrt{\frac{f(D_{OL})}{h(D_{OL})D_{OL}^2-b_c^2f(D_{OL})}}
    \label{eq. rsh}
\end{equation}
In the limit of a distant observer ($D_{OL}\to \infty$), and assuming an asymptotically flat spacetime ($\lim_{r\to \infty}h(r)=1$ and $\lim_{r\to \infty}f(r)=1$), Eq.~\eqref{eq. rsh} becomes
\begin{equation}
    r_{sh} =  b_c,
    \label{eq. shadow rad}
\end{equation}
where the critical impact parameter is given by Eq.~\eqref{eq. crit impact parameter bc}. Since the Kruglov BH is asymptotically flat for $q\neq0$, this relation can be directly applied to compute its shadow radius. For $q\to0^+$, the limit satisfies $\lim_{r\to\infty} h(r)= -1/7$, and therefore Eq.~\eqref{eq. shadow rad} cannot be used in this case.

The BH shadow analysis commonly employs the EHT constraints on the shadow radius of the supermassive Sgr A* BH. From the analysis of Keck and VLTI observational data, Vagnozzi \textit{et al.}~\cite{Vagnozzi2023} derived the following bounds on the dimensionless shadow radius (see also Ref.~\cite{EventHorizonTelescope:2022xqj})
\begin{align}
4.21 < r_{sh}/M < 5.56 \qquad &(2\sigma),\\
4.58 < r_{sh}/M < 5.21 \qquad &(1\sigma).
\end{align}
The resulting shadow radius as a function of $Q$ is shown in Fig.~\ref{rshad}, and the Sgr A* constraints are indicated by the shaded regions.

We first examine the behavior of the shadow radius under variations of $q$, as shown in the left panel of Fig.~\ref{rshad}. A similar pattern is observed to that of the photon sphere radius: for BHs with equivalent charge, small positive (negative) values of $q$ lead to an increase (decrease) in the shadow radius, and it converges to a certain value at large absolute values of $q$. The Sgr A* shadow constraint only allows a particular range of positive $q$ values for BHs with large charge, particularly those with $Q/M\gtrsim0.9$. The constraint also imposes a minimum positive value of $q$ for a given charge, as the shadow radius diverges near $q\to0^+$, although this may be impractical since it still allows extremely small $q$ values\textemdash even for $Q/M=0.98$, the smallest allowed $q$ value under the $1\sigma$ constraint is approximately $q\approx 0.02$.

On the other hand, the right panel of Fig.~\ref{rshad} shows that decreasing positive values of $q$ relax the allowed range of $Q/M$, permitting near-extremal Kruglov BHs with $q=0.1$ and $Q/M\sim1$ to remain consistent with the $2\sigma$ constraint. In contrast, negative values of $q$ yield the opposite effect, tightening the allowed range of $Q/M$ and thereby placing stronger restrictions on Sgr A* being described as a Kruglov BH.

\subsection{Image of accretion disk around the black hole}

\begin{figure*}[htbp!]
	\centering
	\includegraphics[width=1\textwidth]{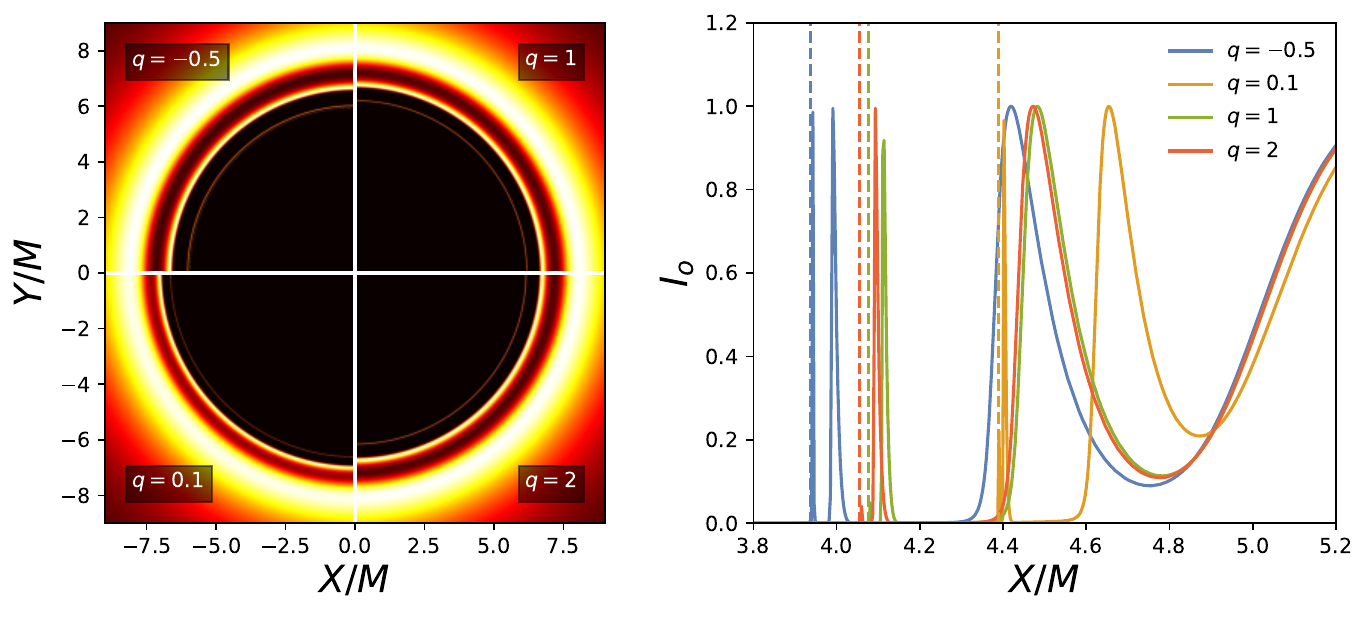}
	\centering
	\caption{(Left) Images of the GLM1 accretion disk surrounding the BH and (Right) its intensity cross section for various values of $q$. The dashed lines in the right panel represent the shadow radius for each configuration. The critical photon ring near the BH shadow becomes wider and more visible for larger positive $q$ or smaller (less negative) values of $q$. The most distinct image appears for $q=0.1$, where the critical photon ring is very thin and the lensed images are significantly shifted outward.}
 \label{shadowimage}
\end{figure*}

\begin{figure*}[htbp!]
	\centering
	\includegraphics[width=1\textwidth]{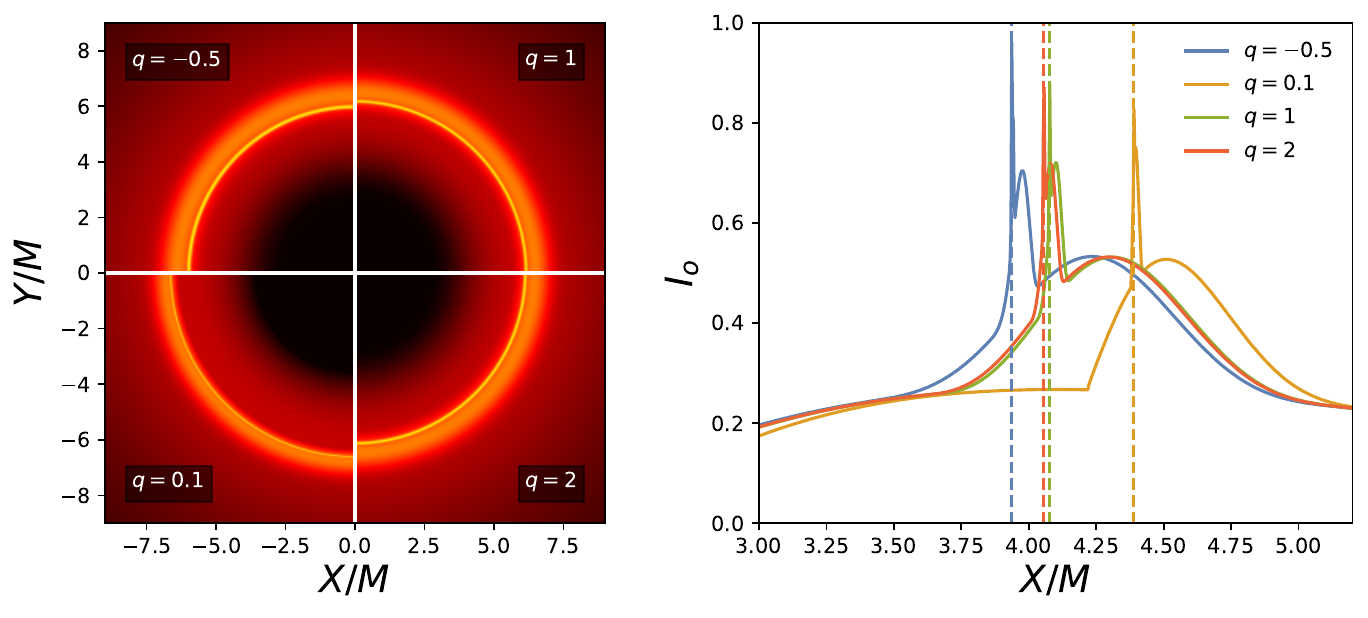}
	\centering
	\caption{(Left) Images of the GLM2 accretion disk surrounding the BH, and (right) its intensity cross section for various values of $q$. The dashed lines in the right panel represent the shadow radius for each configuration. A sharp intensity transition occurs for $q=0.1$ near $X/M\sim4.25$ due to the reversal of the photon’s angular direction.}
 \label{shadowimage-center}
\end{figure*}

We employ a geometrically thin accretion disk with monochromatic emission for simplicity. The emissivity of the disk is described by a simple radially dependent emission profile $I_e(r)$. Specifically, we adopt the Gralla–Lupsasca–Marrone (GLM) model~\cite{Gralla:2020srx}, given by
\begin{equation}
    I_e(r) = \frac{\exp\left\{-\frac{1}{2}\left[\gamma + \operatorname{arcsinh}{\left(\frac{r-\mu}{\sigma}\right)}\right]^2\right\}}{\sqrt{\left(r-\mu\right)^2+\sigma^2}},
    \label{eq. intensity profile glm}
\end{equation}
where the parameters $\gamma$, $\mu$, and $\sigma$ control the radial location, width, and asymmetry of the emission profile, respectively. In contrast to exponential cut-off models (e.g., Refs.~\cite{Meng:2023htc,Zeng:2023fqy,Guo:2022iiy}), the GLM profile provides a continuous intensity function with adjustable smoothness, governed by the parameters $\sigma$ and $\gamma$. This model has been shown to closely align with observational predictions of intensity profiles for astrophysical accretion disks, as derived from general relativistic magnetohydrodynamics~\cite{Vincent:2022fwj}, and has been applied in various studies of BHs and ultracompact objects’ appearances~\cite{Rosa:2023hfm, Rosa:2024bqv, daSilva:2023jxa,Fauzi:2024nta}. We adopt two intensity profiles for the disk:
\begin{enumerate}
    \item \textbf{GLM1}, characterized by the parameters $\gamma = -2$, $\mu = R_{ISCO}$, and $\sigma = M/4$, where $R_{ISCO}$ is the radius of the innermost stable circular orbit (ISCO) for a massive particle. This choice corresponds to an accretion configuration in which the emission stops (and is strongly peaked) near the ISCO. It is particularly useful for studying the features of the photon rings formed around the BH shadow, which are strongly affected by the number of orbits given in Eq.~\eqref{eq. number orbits}. The $R_{ISCO}$ for a spacetime geometry in the form of Eq.~\eqref{line element ori} can be calculated by numerically solving~\cite{Gao:2023mjb}.
    \begin{equation}
    \left.3f(r)f'(r) - 2rf'(r)^2 + rf(r)f''(r)\right|_{r=R_{ISCO}} = 0.
    \end{equation}

    \item \textbf{GLM2}, characterized by the parameters $\gamma = \mu = 0$ and $\sigma = 2M$. This accretion profile extends up to the center $r=0$ without any cut-off at the ISCO or the horizon. This type of accretion disk may be irrelevant for BH observations; however, as seen in Fig.~\ref{Shadowtrace}, there is a unique frame-dragging-like effect occurring near the BH horizon. Therefore, we use this particular disk profile, which spans all the way to the horizon, to probe its effect on the BH image.
\end{enumerate}

We limit our BH imaging to axial observation, in which the observer’s line of sight is perpendicular to the plane of the accretion disk. This is particularly useful for identifying the discrepancy between BHs described by Maxwell electrodynamics and others under ideal conditions. It is also sufficient to assume a static accretion disk, which allows us to neglect the Doppler redshift effect, as the motion of particles in the accretion disk has no relative difference on either side. Therefore, we only consider the gravitational redshift, which simply reduces the observed intensity $I_o$ by
\begin{equation}
    I_o=A(r)^2I_e.
\end{equation}

In Fig.~\ref{shadowimage}, we present a comparison of the simulated images of a near-extremal Kruglov BH with $Q/M=0.98$ for several values of $q$, along with their intensity cross sections using the GLM1 accretion disk intensity profile. Asymptotically, they all share identical direct images of the accretion disk, independent of $q$. The differences occur predominantly in the strongly lensed images and, most notably, in the structure of the inner photon ring. The location of the inner photon ring can be interpreted as the boundary of the BH shadow radius. This interpretation is corroborated by the intensity profiles shown in the right panel of Fig.~\ref{shadowimage}, where the innermost intensity spikes (corresponding to the inner photon ring) consistently appear just outside the shadow radius (indicated by the dashed lines).

For $q=0.1$, the lensed image of the accretion disk is shifted to larger angular radii and appears noticeably narrower compared to the other configurations. In this case, the inner photon ring is effectively suppressed and remains unresolved at the present imaging resolution, suggesting that substantially higher angular resolution would be required to clearly detect it. In contrast, the case $q=-0.5$ produces a broader lensed image that is shifted inward, as well as greater visibility of its inner photon ring feature. This behavior is consistent with our earlier discussion in Sec.~\ref{sec:deflection} about the range of impact parameters that give rise to photon-ring trajectories. We therefore conclude that smaller positive values of $q$ generically suppress the observability of lensed accretion disk images.

The accretion disk image around the BH with the GLM2 intensity profile is shown in Fig.~\ref{shadowimage-center}. The overall features are largely similar to those of the GLM1 profile; however, a distinct new feature appears for the case $q=0.1$. In this case, not only do the photon ring and lensed images shift outward, but a sharp intensity fall-off also emerges near $X/M \sim 4.25$ in the intensity cross section. For other values of $q$, the transition remains smooth. This behavior is a direct consequence of the change in the angular direction of photon trajectories for sufficiently small $q$, as discussed in Sec.~\ref{sec. null geodesic} and illustrated in Fig.~\ref{Shadowtrace}. Specifically, some photons with $b < b_c$ reverse their direction before reaching the equatorial plane where the accretion disk is located. Meanwhile, other photons with slightly different impact parameters do reach the equatorial plane, then reverse their angular direction, and intersect the plane again, thereby amplifying the observed intensity. This mechanism produces the sharp feature in the intensity profile and may serve as a signature of the presence of a photon angular turning point.

\section{Conclusion}
\label{sec:conc}

The modification of Maxwell electrodynamics through nonlinear electrodynamics (NLED) has long been motivated by both conceptual and phenomenological considerations. One of the earliest realizations is the Born–Infeld (BI) theory, which extends the Maxwell Lagrangian by incorporating nonlinear corrections that regularize the electromagnetic field at high energies. The Kruglov model can be viewed as a natural generalization of this framework. Instead of a fixed square-root structure as in BI theory, it introduces a one-parameter family of Lagrangians governed by the exponent $q$. This formulation interpolates smoothly between Maxwell electrodynamics ($q=1$) and BI theory ($q=1/2$), while remaining well-defined for arbitrary values of $q$. 

An appealing feature of this model is that, for sufficiently small $\beta$, the spacetime geometry outside the event horizon remains very close to the Reissner--Nordström solution, whereas the effective geometry governing photon propagation can differ substantially. This separation allows the optical consequences of nonlinear electrodynamics to be investigated largely independently of modifications to the background spacetime itself.

In this work, we investigated the effective null geodesic structure associated with magnetically charged black holes in Kruglov electrodynamics and analyzed its observational consequences through fully numerical calculations. Rather than relying on strong-field analytical approximations, photon trajectories were obtained by direct numerical integration, enabling the effective geometry to be explored over the entire range of relevant impact parameters. We showed that the nonlinear parameter $q$ governs the qualitative structure of the photon effective potential and, consequently, the behavior of null geodesics. In particular, sufficiently small positive values of $q$ generate a pole in the effective optical geometry, leading to the appearance of stable photon orbits outside the event horizon and causing the azimuthal evolution of photon trajectories to reverse locally near the black hole, producing trajectory patterns that resemble, although do not originate from, frame-dragging effects in rotating spacetimes. Conversely, negative values of $q$ produce a zero of the effective correction factor, which acts as a turning point for photon trajectories and prevents photons from entering the region where the background metric becomes complex. These features arise solely from the effective optical geometry and do not correspond to curvature singularities of the underlying spacetime.


The deformation of the effective photon geometry produces systematic modifications of all optical observables considered in this work. Smaller positive values of $q$ enhance the gravitational deflection of light, increase the characteristic scales associated with the unstable photon orbit, and enlarge the shadow radius, whereas negative values of $q$ generally lead to the opposite behavior. Using the observational constraints on the Sgr~A* shadow radius inferred from the comprehensive EHT analysis of Ref.~\cite{Vagnozzi2023}, we found that a significant region of the Kruglov parameter space remains compatible with current observations. Although positive values of $q$ relax the allowed charge-to-mass ratio compared with the Maxwell case, the present observational uncertainties remain too large to place meaningful constraints on the nonlinear parameter itself.

Based on the effective photon geometry, we also investigated the appearance of a thin accretion disk using a simplified static, optically thin, monochromatic emission model described by the GLM emissivity profile. The purpose of adopting this simplified setup is to isolate the influence of the effective optical geometry on image formation while minimizing astrophysical uncertainties. We found that the morphology of the photon ring depends sensitively on the nonlinear parameter through its modification of the effective null structure. Smaller positive values of $q$ reduce the range of impact parameters supporting multiple photon trajectories, producing thinner and less prominent photon rings, whereas negative values of $q$ broaden the corresponding ring structures. These results demonstrate that nonlinear electrodynamics can produce observable modifications to black-hole images even when the underlying spacetime geometry remains nearly indistinguishable from the corresponding RN solution.

Several aspects of the present analysis deserve further investigation. The stable photon orbits identified here are defined within the geometric-optics approximation through extrema of the effective potential. A complete assessment of their physical significance requires analysis beyond the present framework, including wave propagation, perturbative stability of the effective optical geometry, hyperbolicity and causality of the underlying nonlinear electrodynamics, as well as realistic general relativistic magnetohydrodynamic (GRMHD) models of black-hole accretion flows~\cite{EventHorizonTelescope:2022urf, Vincent:2022fwj}. Such investigations would clarify the robustness of the optical signatures predicted here under more realistic astrophysical conditions.

Overall, the Kruglov model provides a particularly useful theoretical laboratory for disentangling modifications of photon propagation from those arising directly from the spacetime geometry. The present results demonstrate that effective optical geometries alone can generate substantial and potentially observable signatures in strong gravitational lensing, black-hole shadows, and horizon-scale imaging. This highlights the importance of effective photon geometries as an independent probe of NLED in the strong-field regime and motivates further studies of their implications for future high-resolution black-hole observations.



\section*{Acknowledgements}

We thank Byon Jayawiguna, Faris Darmawan, Fernanda Pratama, and Iffatricia Haura for the valuable discussions. This work is supported by Hibah PUTI Q1 UI No.~PKS-196/UN2.RST/HKP.05.00/2025.

\section*{Declaration of generative AI and AI-assisted technologies in the manuscript preparation process}

During the preparation of this work the authors used ChatGPT in order to improve language, sentence flow, readability, and overall clarity. After using this tool, the authors reviewed and edited the content as needed and take full responsibility for the content of the published article.

\begin{appendix}
\section{Comparison with the strong field limit approximation}
\label{sec. strong field approx}

Here, we briefly show the method originally proposed by Bozza in Ref.~\cite{Bozza2002} and extended by Tsukamoto in Ref.~\cite{Tsukamoto2017} to calculate the approximate deflection angle in the strong field limit. For our convenience, let us re-write the line element as
\begin{equation}
    ds^2 = -A(r)dt^2 + B(r)dr^2 + C(r)d\Omega^2,
    \label{eq. SSS AB}
\end{equation}
which generalizes the effective line element in Eq.~\eqref{effline}, where
\begin{equation}
        A(r) = \frac{1}{B(r)} = f(r),\qquad C(r) = h(r)r^2.
\end{equation}

\begin{figure}[htbp]
	\centering
\includegraphics[width=0.7\textwidth]{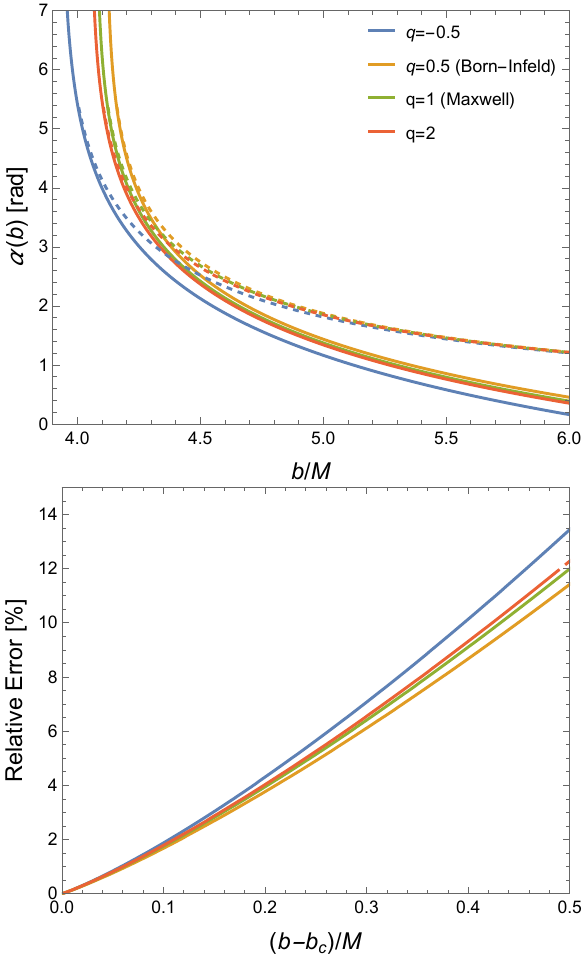}
	\centering
	\caption{(Top) Deflection angle as a function of the impact parameter obtained by strong field approximation (solid lines) and full numerical procedure (dashed lines). (Bottom) Relative error between the strong field approximation and the full numerical approach, following the relation $|\alpha_N-\alpha_A|/\alpha_A$ where $\alpha_N$ and $\alpha_A$ is the full numerical and approximation results of the deflection angle, respectively.}
 \label{fig. numerical vs approx deflection angle}
\end{figure}

We define the integration typically used in the approximation,
\begin{equation}
    T(r_0) = 2 \int^{\infty}_{r_0} \left(\frac{d\phi}{dr}\right) dr,
    \label{T}
\end{equation}
which is equivalent to $\phi_{end}$ in Eq.~\eqref{eq. phi end} in the limit $D_{OL}=D_{LS}\to\infty$. Introducing a new variable $y=1-r_0/r$, we redefine $T(r_0)$ into
\begin{align}
    T(r_0)=\int^{1}_{0} \mathcal{H}(y,r_0) dy,\label{fj}\\
    \qquad \mathcal{H}(y,r_0)\equiv\frac{2r_0}{\sqrt{G(y,r_0)}},\notag
\end{align}
and the $G(y, r_0)$ is given by
\begin{align}
    G(y, r_0)&=\frac{R(r_0)C(r_0)}{B(r_0)}(1 - y)^4,\label{ohiho}\\
    R(r) &= \frac{C(r)}{A(r)b^2} - 1. \notag
\end{align}
This function can be expanded near $y=0$ into
\begin{equation}
    G(y,r_0) =  \sum_{n=1}^{\infty} \mathcal{N}_n(r_0)y^n,
    \label{jan}
\end{equation}
and the expansion term up to the second order is given by
\begin{eqnarray}
    \mathcal{N}_1(r_0) &=& \frac{C(r_0)D(r_0)r_0}{B(r_0)},\\
    \mathcal{N}_2(r_0) &=& \frac{C(r_0)r_0}{B(r_0)}\left\{D(r_0) \left[\left(D(r_0) - \frac{B'(r_0)}{B(r_0)} \right)r_0 - 3 \right] + \frac{r_0}{2}\left[\frac{C''(r_0)}{C(r_0)}- \frac{A''(r_0)}{A(r_0)}\right] \right\}.
\end{eqnarray}

Following Tsukamoto procedure, the integral can be written as the divergent (D) and the regular (R) parts. It is expressed as
\begin{equation}
    \mathcal{H}(y,r_0)=\mathcal{H}_D(y,r_0)+\mathcal{H}_R(y,r_0).
    \label{Reg}
\end{equation}
Using the expansion of Eq.~\eqref{jan} up to the second order, the divergent part of Eq.~\eqref{fj} reads
\begin{equation}
    \mathcal{H}_D(y, r_0) = \frac{2r_0}{\sqrt{\mathcal{N}_1(r_0)y + \mathcal{N}_2(r_0)y^2}}.
    \label{divvv}
\end{equation}
However, at strong field limit $r_0\to r_{ps}$, one have $D(r_0)=0$ and hence $\mathcal{N}_1 = 0$, therefore Eq.~\eqref{divvv} becomes
\begin{align}
    \lim_{r_0\to r_{ps}}\mathcal{H}_D(y, r_{0}) =&\lim_{r_0\to r_{ps}}\frac{2r_{0}}{\sqrt{\mathcal{N}_2(r_{0})y^2}},\\
    \lim_{r_0\to r_{ps}}\mathcal{N}_2(r_0)=& \frac{C(r_0)r_0^2}{2B(r_0)}\left[\frac{C''(r_0)}{C(r_0)}- \frac{A''(r_0)}{A(r_0)}\right].
\end{align}

Now, the regular part can be obtain via Eq.~\eqref{Reg}, and its integral is
\begin{equation}
    T_R(r_0) \equiv \int^1_0 \left[\mathcal{H}(y,r_0)-\mathcal{H}_D(y,r_0)\right] dy.
\end{equation}
This integral will be used to obtain the deflection angle $\alpha$, specifically to obtain the coefficient $\tilde{a}_2$. The equation for deflection angle in the strong field limit is
\begin{equation}
    \alpha(b) = -\tilde{a}_1 \log \left(\frac{\tilde{b}}{b_c} \right) + \tilde{a}_2 + \mathcal{O}\left[\tilde{b}\log\tilde{b}\right],
    \label{eq. deflection alpha A}
\end{equation}
where $\tilde{b}=b-b_c$, while the coefficient $\tilde{a}_1$ and $\tilde{a}_2$ is given by
\begin{align}
    \tilde{a}_1 =& \sqrt{\frac{2B(r_{ps})A(r_{ps})}{C''(r_{ps})A(r_{ps}) - C(r_{ps})A''(r_{ps})}},\\
    \tilde{a}_2 =& \tilde{a}_1 \log \left\{r^{2}_{ps} \left[\frac{C''(r_{ps})}{C(r_{ps})}-\frac{A''(r_{ps})}{A(r_{ps})}\right] \right\}+ T_R(r_{ps}) - \pi.
\end{align}
Obtaining the two coefficients $\tilde{a}_1$ and $\tilde{a}_2$ allows us to analytically determine the deflection angle for a given impact parameter by using Eq.~\eqref{eq. deflection alpha A}.

We compare the fully numerical and the approximate deflection angle results together with its relative error in Fig.~\ref{fig. numerical vs approx deflection angle} for representative values of $q$ and $Q/M=0.98$. We use the expanded metric in the form of Eq.~\eqref{expandf} to avoid the complex integration, hence the chosen values of $q$ are those that well fit with the exact form.

The approximate results agree with the numerical computations in the immediate vicinity of the critical impact parameter. However, the approximation error becomes non-negligible once the impact parameter exceeds $b_c$ just by a slightly higher amount. For example, even at $(b - b_c)/M \sim 0.5$, the relative error exceeds $10\%$ compared to the full numerical value. Furthermore, at larger impact parameters, the approximation becomes unreliable, as it exhibits an unphysical monotonic decrease and eventually attains negative values. We therefore do not rely on the strong-field limit approximation in our analysis, as it may lead to misleading interpretations in the present context.

\end{appendix}


\begin{thebibliography}{99}


\bibitem{Schneider:1992bmb}
P.~Schneider, J.~Ehlers and E.~E.~Falco,
``Gravitational Lenses,''
\Journal{}{Springer, 1992}
{10.1007/978-3-662-03758-4}.

\bibitem{Schneider2}
P.~Schneider, C.~S.~Kochanek, and J.~Wambsganss, “Introduction to gravitational lensing and cosmology,” in
\Journal{Gravitational Lensing: Strong, Weak and Micro}{(2006)}{10.1007/978-3-540-30310-7_1}.

\bibitem{DE1919}
F.~W.~Dyson, A.~S.~Eddington and C.~Davidson,
``A Determination of the Deflection of Light by the Sun's Gravitational Field, from Observations Made at the Total Eclipse of May 29, 1919,''
\Journal{Phil. Trans. Roy. Soc. Lond. A}{\textbf{220}, 291 (1920)}
{10.1098/rsta.1920.0009}.

\bibitem{Lodge}
O.~J.~Lodge, ``Gravitation and Light," 
\Journal{Nature}{\textbf{104}, 354 (1919)}{10.1038/104354a0}.

\bibitem{Liebes:1964zz}
S.~Liebes,
``Gravitational Lenses,''
\Journal{Phys. Rev.}{\textbf{133}, B835 (1964)}
{10.1103/PhysRev.133.B835}.

\bibitem{Refsdal:1964yk}
S.~Refsdal,
``The gravitational lens effect,''
\Journal{Mon. Not. Roy. Astron. Soc.}{\textbf{128}, 295 (1964)}{10.1093/mnras/128.4.295}.

\bibitem{Blandford:1991xc}
R.~D.~Blandford and R.~Narayan,
``Cosmological applications of gravitational lensing,''
\Journal{Ann. Rev. Astron. Astrophys.}{\textbf{30}, 311 (1992)}
{10.1146/annurev.aa.30.090192.001523}.

\bibitem{Perlick:2004tq}
V.~Perlick,
``Gravitational lensing from a spacetime perspective,''
\Journal{Living Rev. Rel.}{\textbf{7}, 9 (2004)}
{10.12942/lrr-2004-9}.

\bibitem{darwin}
C.~Darwin, 
``The gravity field of a particle I,''
\Journal{Proc. Roy. Soc. Lond.}{\textbf{A} 249, 180 (1959)}{10.1098/rspa.1959.0015}.

\bibitem{Bozza2010}
V.~Bozza,
``Gravitational Lensing by Black Holes,''
\Journal{Gen. Rel. Grav.}{\textbf{42}, 2269 (2010)}
{10.1007/s10714-010-0988-2},
\arXiv{0911.2187}{gr-qc}.

\bibitem{Luminet:1979nyg}
J.~P.~Luminet,
``Image of a spherical black hole with thin accretion disk,''
Astron. Astrophys. \textbf{75}, 228 (1979).

\bibitem{Falcke:1999pj}
H.~Falcke, F.~Melia and E.~Agol,
``Viewing the shadow of the black hole at the galactic center,''
\Journal{Astrophys. J. Lett.}{\textbf{528} (2000), L13}
{10.1086/312423},
\arXiv{astro-ph/9912263}{astro-ph}.

\bibitem{EventHorizonTelescope:2019dse}
K.~Akiyama \textit{et al.} [Event Horizon Telescope],
``First M87 Event Horizon Telescope Results. I. The Shadow of the Supermassive Black Hole,''
\Journal{Astrophys. J. Lett.}{\textbf{875}, L1 (2019)}
{10.3847/2041-8213/ab0ec7},
\arXiv{1906.11238}{astro-ph.GA}.

\bibitem{EventHorizonTelescope:2019uob}
K.~Akiyama \textit{et al.} [Event Horizon Telescope],
``First M87 Event Horizon Telescope Results. II. Array and Instrumentation,''
\Journal{Astrophys. J. Lett.}{\textbf{875}, L2 (2019)}
{10.3847/2041-8213/ab0c96},
\arXiv{1906.11239}{astro-ph.IM}.

\bibitem{EventHorizonTelescope:2019jan}
K.~Akiyama \textit{et al.} [Event Horizon Telescope],
``First M87 Event Horizon Telescope Results. III. Data Processing and Calibration,''
\Journal{Astrophys. J. Lett.}{\textbf{875}, L3 (2019)}
{10.3847/2041-8213/ab0c57},
\arXiv{1906.11240}{astro-ph.GA}.

\bibitem{EventHorizonTelescope:2019ths}
K.~Akiyama \textit{et al.} [Event Horizon Telescope],
``First M87 Event Horizon Telescope Results. IV. Imaging the Central Supermassive Black Hole,''
\Journal{Astrophys. J. Lett.}{\textbf{875}, L4 (2019)}
{10.3847/2041-8213/ab0e85},
\arXiv{1906.11241}{astro-ph.GA}.

\bibitem{EventHorizonTelescope:2019pgp}
K.~Akiyama \textit{et al.} [Event Horizon Telescope],
``First M87 Event Horizon Telescope Results. V. Physical Origin of the Asymmetric Ring,''
\Journal{Astrophys. J. Lett.}{\textbf{875}, L5 (2019)}
{10.3847/2041-8213/ab0f43},
\arXiv{1906.11242}{astro-ph.GA}.

\bibitem{EventHorizonTelescope:2019ggy}
K.~Akiyama \textit{et al.} [Event Horizon Telescope],
``First M87 Event Horizon Telescope Results. VI. The Shadow and Mass of the Central Black Hole,''
\Journal{Astrophys. J. Lett.}{\textbf{875}, L6 (2019)}
{10.3847/2041-8213/ab1141},
\arXiv{1906.11243}{astro-ph.GA}.

\bibitem{EventHorizonTelescope:2021bee}
K.~Akiyama \textit{et al.} [Event Horizon Telescope],
``First M87 Event Horizon Telescope Results. VII. Polarization of the Ring,''
\Journal{Astrophys. J. Lett.}{\textbf{910}, L12 (2021)}
{10.3847/2041-8213/abe71d},
\arXiv{2105.01169}{astro-ph.HE}.

\bibitem{EventHorizonTelescope:2021srq}
K.~Akiyama \textit{et al.} [Event Horizon Telescope],
``First M87 Event Horizon Telescope Results. VIII. Magnetic Field Structure near The Event Horizon,''
\Journal{Astrophys. J. Lett.}{\textbf{910}, L13 (2021)}
{10.3847/2041-8213/abe4de},
\arXiv{2105.01173}{astro-ph.HE}.

\bibitem{EventHorizonTelescope:2023gtd}
K.~Akiyama \textit{et al.} [Event Horizon Telescope],
``First M87 Event Horizon Telescope Results. IX. Detection of Near-horizon Circular Polarization,''
\Journal{Astrophys. J. Lett.}{\textbf{957}, L20 (2023)}
{10.3847/2041-8213/acff70},
\arXiv{2311.10976}{astro-ph.HE}.

\bibitem{EventHorizonTelescope:2022wkp}
K.~Akiyama \textit{et al.} [Event Horizon Telescope],
``First Sagittarius A* Event Horizon Telescope Results. I. The Shadow of the Supermassive Black Hole in the Center of the Milky Way,''
\Journal{Astrophys. J. Lett.}{\textbf{930}, L12 (2022)}
{10.3847/2041-8213/ac6674}
\arXiv{2311.08680}{astro-ph.HE}.

\bibitem{EventHorizonTelescope:2022apq}
K.~Akiyama \textit{et al.} [Event Horizon Telescope],
``First Sagittarius A* Event Horizon Telescope Results. II. EHT and Multiwavelength Observations, Data Processing, and Calibration,''
\Journal{Astrophys. J. Lett.}{\textbf{930}, L13 (2022)}
{10.3847/2041-8213/ac6675},
\arXiv{2311.08679}{astro-ph.HE}.

\bibitem{EventHorizonTelescope:2022wok}
K.~Akiyama \textit{et al.} [Event Horizon Telescope],
``First Sagittarius A* Event Horizon Telescope Results. III. Imaging of the Galactic Center Supermassive Black Hole,''
\Journal{Astrophys. J. Lett.}{\textbf{930}, L14 (2022)}
{10.3847/2041-8213/ac6429},
\arXiv{2311.09479}{astro-ph.HE}.

\bibitem{EventHorizonTelescope:2022exc}
K.~Akiyama \textit{et al.} [Event Horizon Telescope],
``First Sagittarius A* Event Horizon Telescope Results. IV. Variability, Morphology, and Black Hole Mass,''
\Journal{Astrophys. J. Lett.}{\textbf{930}, L15 (2022)}
{10.3847/2041-8213/ac6736},
\arXiv{2311.08697}{astro-ph.HE}.

\bibitem{EventHorizonTelescope:2022urf}
K.~Akiyama \textit{et al.} [Event Horizon Telescope],
``First Sagittarius A* Event Horizon Telescope Results. V. Testing Astrophysical Models of the Galactic Center Black Hole,''
\Journal{Astrophys. J. Lett.}{\textbf{930}, L16 (2022)}
{10.3847/2041-8213/ac6672},
\arXiv{2311.09478}{astro-ph.HE}.

\bibitem{EventHorizonTelescope:2022xqj}
K.~Akiyama \textit{et al.} [Event Horizon Telescope],
``First Sagittarius A* Event Horizon Telescope Results. VI. Testing the Black Hole Metric,''
\Journal{Astrophys. J. Lett.}{\textbf{930}, L17 (2022)}
{10.3847/2041-8213/ac6756},
\arXiv{2311.09484}{astro-ph.HE}.

\bibitem{EventHorizonTelescope:2024hpu}
K.~Akiyama \textit{et al.} [Event Horizon Telescope],
``First Sagittarius A* Event Horizon Telescope Results. VII. Polarization of the Ring,''
\Journal{Astrophys. J. Lett.}{\textbf{964}, L25 (2024)}
{10.3847/2041-8213/ad2df0}.

\bibitem{EventHorizonTelescope:2024rju}
K.~Akiyama \textit{et al.} [Event Horizon Telescope],
``First Sagittarius A* Event Horizon Telescope Results. VIII. Physical Interpretation of the Polarized Ring,''
\Journal{Astrophys. J. Lett.}{\textbf{964}, L26 (2024)}
{10.3847/2041-8213/ad2df1}.

\bibitem{Boos:2025nzc}
J.~Boos and H.~Hu,
``Microlensing of nonsingular black holes at finite size: A ray tracing approach,''
\Journal{Phys. Rev. D}{\textbf{113} (2026) no.2, 024065}
{10.1103/dv28-xm9x},
\arXiv{2510.10282}{gr-qc}.

\bibitem{bardeen}
J.~Bardeen, ``Non-singular general-relativistic gravitational collapse,'' \textit{Proc.~Int.~Conf.~GR5,} Tbilisi \textbf{174}, 1968.

\bibitem{Ayon-Beato:2000mjt}
E.~Ayon-Beato and A.~Garcia,
``The Bardeen model as a nonlinear magnetic monopole,''
\Journal{Phys. Lett. B}{\textbf{493}, 149 (2000)}
{10.1016/S0370-2693(00)01125-4},
\arXiv{gr-qc/0009077}{gr-qc}.

\bibitem{BI1934}
M.~Born and L.~Infeld,
``Foundations of the new field theory,''
\Journal{Nature}{\textbf{132}, 1004.1 (1933)}
{10.1038/1321004b0}.

\bibitem{Pellicer:1969cf}
R.~Pellicer and R.~J.~Torrence,
``Nonlinear electrodynamics and general relativity,''
\Journal{J. Math. Phys.}{\textbf{10}, 1718 (1969)}
{10.1063/1.1665019}.

\bibitem{Salazar:1987ap}
I.~H.~Salazar, A.~Garcia and J.~Plebanski,
``Duality Rotations and Type $D$ Solutions to Einstein Equations With Nonlinear Electromagnetic Sources,''
\Journal{J. Math. Phys.}{\textbf{28}, 2171 (1987)}
{10.1063/1.527430}.

\bibitem{Breton2010}
N.~Breton and R.~Garcia-Salcedo,
``Nonlinear Electrodynamics and black holes,''
\arXiv{hep-th/0702008}{hep-th}.

\bibitem{Tommasini:2008lrs}
D.~Tommasini, A.~Ferrando, H.~Michinel and M.~Seco,
``Detecting photon-photon scattering in vacuum at exawatt lasers,''
\Journal{Phys. Rev. A}{\textbf{77}, 042101 (2008)}
{10.1103/PhysRevA.77.042101},
\arXiv{0802.0101}{physics.optics}.

\bibitem{Pike:2014wha}
O.~J.~Pike, F.~Mackenroth, E.~G.~Hill and S.~J.~Rose,
``A photon{\textendash}photon collider in a vacuum hohlraum,''
\Journal{Nature Photon}{\textbf{8}, 434 (2014)}
{10.1038/nphoton.2014.95}.

\bibitem{Gaete:2014nda}
P.~Gaete and J.~Helay{\"e}l-Neto,
``Remarks on nonlinear Electrodynamics,''
\Journal{Eur. Phys. J. C}{\textbf{74}, 3182 (2014)}
{10.1140/epjc/s10052-014-3182-y},
\arXiv{1408.3363}{hep-th}.

\bibitem{Gaete:2022lkf}
P.~Gaete and J.~A.~Helay{\"e}l-Neto,
``Vacuum material properties and Cherenkov radiation in logarithmic electrodynamics,''
\Journal{Eur. Phys. J. C}{\textbf{83}, 128 (2023)}
{10.1140/epjc/s10052-023-11280-w},
\arXiv{2205.03252}{hep-ph}.

\bibitem{Kadlecova:2023qxn}
H.~Kadlecova,
``Photon-photon scattering in Born-Infeld electrodynamics,''
\Journal{Proc. SPIE Int. Soc. Opt. Eng.}{\textbf{12580}, 1258007 (2023)}
{10.1117/12.2665647}.

\bibitem{2Kruglov2017}
S.~I.~Kruglov,
``Notes on Born\textendash{}Infeld-type electrodynamics,''
\Journal{Mod. Phys. Lett. A}{\textbf{32}, 1750201 (2017)}
{10.1142/S0217732317502017},
\arXiv{1612.04195}{physics.gen-ph}.

\bibitem{Kruglov2017}
S.~I.~Kruglov,
``Born\textendash{}Infeld-type electrodynamics and magnetic black holes,''
\Journal{Annals Phys.}{\textbf{383}, 550 (2017)}
{10.1016/j.aop.2017.06.008},
\arXiv{1707.04495}{gr-qc}.

\bibitem{Russo:2024kto}
J.~G.~Russo and P.~K.~Townsend,
``Born again,''
\Journal{SciPost Phys.}{\textbf{16}, 124 (2024)}
{10.21468/SciPostPhys.16.5.124},
\arXiv{2401.04167}{hep-th}.

\bibitem{Russo:2026vnj}
J.~G.~Russo and P.~K.~Townsend,
``Black holes and causal nonlinear electrodynamics,''
\arXiv{2601.07789}{hep-th}.

\bibitem{Novello2000}
M.~Novello, V.~A.~De Lorenci, J.~M.~Salim and R.~Klippert,
``Geometrical aspects of light propagation in nonlinear electrodynamics,''
\Journal{Phys. Rev. D}{\textbf{61}, 045001 (2000)}
{10.1103/PhysRevD.61.045001},
\arXiv{gr-qc/9911085}{gr-qc}.

\bibitem{AH2020}
A.~S.~Habibina and H.~S.~Ramadhan,
``Geodesic of nonlinear electrodynamics and stable photon orbits,''
\Journal{Phys. Rev. D}{\textbf{101}, 124036 (2020)}
{10.1103/PhysRevD.101.124036},
\arXiv{2007.03211}{gr-qc}.

\bibitem{HSR2023}
H.~S.~Ramadhan, M.~F.~Ishlah, F.~P.~Pratama and I.~Alfredo,
``Strong lensing and shadow of Ayon-Beato\textendash{}Garcia (ABG) nonsingular black hole,''
\Journal{Eur. Phys. J. C}{\textbf{83}, 465 (2023)}
{10.1140/epjc/s10052-023-11648-y},
\arXiv{2303.10921}{gr-qc}.

\bibitem{Fauzi:2025asu}
M.~F.~Fauzi,
``Comment on {\textquotedblleft}Strong lensing and shadow of Ayon-Beato{\textendash}Garcia (ABG) nonsingular black hole{\textquotedblright},''
\Journal{Eur. Phys. J. C}{\textbf{85}, 1246 (2025)}
{10.1140/epjc/s10052-025-14991-4},
\arXiv{2509.24777}{gr-qc}.

\bibitem{Tlemissov:2025nnk}
A.~Tlemissov, B.~Toshmatov and J.~Kov{\'a}{\v{r}},
``Effect of nonlinear electrodynamics on polarization distribution around black holes,''
\Journal{Phys. Rev. D}{\textbf{111}, 064084 (2025)}
{10.1103/PhysRevD.111.064084},
\arXiv{2503.08294}{gr-qc}.

\bibitem{KumarWalia:2024yxn}
R.~Kumar Walia,
``Exploring nonlinear electrodynamics theories: Shadows of regular black holes and horizonless ultracompact objects,''
\Journal{Phys. Rev. D}{\textbf{110}, 064058 (2024)}
{10.1103/PhysRevD.110.064058},
\arXiv{2409.13290}{gr-qc}.

\bibitem{Guzman-Herrera:2024fkg}
E.~Guzman-Herrera, A.~Montiel and N.~Breton,
``Comparative of light propagation in Born-Infeld, Euler-Heisenberg and ModMax nonlinear electrodynamics,''
\Journal{JCAP}{\textbf{11}, 002 (2024)}
{10.1088/1475-7516/2024/11/002},
\arXiv{2407.21326}{gr-qc}.

\bibitem{Murk:2024nod}
S.~Murk and I.~Soranidis,
``Light rings and causality for nonsingular ultracompact objects sourced by nonlinear electrodynamics,''
\Journal{Phys. Rev. D}{\textbf{110}, 044064 (2024)}
{10.1103/PhysRevD.110.044064},
\arXiv{2406.07957}{gr-qc}.

\bibitem{Bronnikov:2000vy}
K.~A.~Bronnikov,
``Regular magnetic black holes and monopoles from nonlinear electrodynamics,''
\Journal{Phys. Rev. D}{\textbf{63}, 044005 (2001)}
{10.1103/PhysRevD.63.044005},
\arXiv{gr-qc/0006014}{gr-qc}.

\bibitem{Bronnikov:2017sgg}
K.~A.~Bronnikov,
``Nonlinear electrodynamics, regular black holes and wormholes,''
\Journal{Int. J. Mod. Phys. D}{\textbf{27}, 1841005 (2018)}
{10.1142/S0218271818410055},
\arXiv{1711.00087}{gr-qc}.

\bibitem{Flores-Alfonso:2020euz}
D.~Flores-Alfonso, B.~A.~Gonz{\'a}lez-Morales, R.~Linares and M.~Maceda,
``Black holes and gravitational waves sourced by non-linear duality rotation-invariant conformal electromagnetic matter,''
\Journal{Phys. Lett. B}{\textbf{812}, 136011 (2021)}
{10.1016/j.physletb.2020.136011},
\arXiv{2011.10836}{gr-qc}.

\bibitem{Hendi:2013mka}
S.~H.~Hendi and A.~Sheykhi,
``Charged rotating black string in gravitating nonlinear electromagnetic fields,''
\Journal{Phys. Rev. D}{\textbf{88} (2013) no.4, 044044}
{10.1103/PhysRevD.88.044044},
\arXiv{1405.6998}{gr-qc}.

\bibitem{Claudel:2000yi}
C.~M.~Claudel, K.~S.~Virbhadra and G.~F.~R.~Ellis,
``The Geometry of photon surfaces,''
\Journal{J. Math. Phys.}{\textbf{42}, 818 (2001)}
{10.1063/1.1308507},
\arXiv{gr-qc/0005050}{gr-qc}.

\bibitem{Cardoso:2008bp}
V.~Cardoso, A.~S.~Miranda, E.~Berti, H.~Witek and V.~T.~Zanchin,
``Geodesic stability, Lyapunov exponents and quasinormal modes,''
\Journal{Phys. Rev. D}{\textbf{79}, 064016 (2009)}
{10.1103/PhysRevD.79.064016},
\arXiv{0812.1806}{hep-th}.

\bibitem{Synge:1966okc}
J.~L.~Synge,
``The Escape of Photons from Gravitationally Intense Stars,''
\Journal{Mon. Not. Roy. Astron. Soc.}{\textbf{131}, 463 (1966)}
{10.1093/mnras/131.3.463}.

\bibitem{Bardeen:1973tla}
J.~M.~Bardeen,
``Timelike and null geodesics in the Kerr metric,''
Proceedings, Ecole d'Et{\'e} de Physique Th{\'e}orique: Les Astres Occlus : Les Houches, France, August, 1972, 215 (1973)

\bibitem{Hod:2013jhd}
S.~Hod,
``Upper bound on the radii of black-hole photonspheres,''
\Journal{Phys. Lett. B}{\textbf{727}, 345 (2013)}
{10.1016/j.physletb.2013.10.047},
\arXiv{1701.06587}{gr-qc}.

\bibitem{Bozza2002}
V.~Bozza,
``Gravitational lensing in the strong field limit,''
\Journal{Phys. Rev. D}{ \textbf{66}, 103001 (2002)}
{10.1103/PhysRevD.66.103001},
\arXiv{gr-qc/0208075}{gr-qc}.

\bibitem{Tsukamoto2017}
N.~Tsukamoto,
``Deflection angle in the strong deflection limit in a general asymptotically flat, static, spherically symmetric spacetime,''
\Journal{Phys. Rev. D}{\textbf{95}, 064035 (2017)}
{10.1103/PhysRevD.95.064035},
\arXiv{1612.08251}{gr-qc}.

\bibitem{Pereira:2025fvg}
C.~F.~S.~Pereira, A.~R.~Soares, M.~V.~d.~S.~Silva, R.~L.~L.~Vit{\'o}ria and H.~Belich,
``Light deflection and gravitational lensing effects in acoustic black-bounce spacetime,''
\Journal{Phys. Rev. D}{\textbf{112}, 064012 (2025)}
{10.1103/fqvd-8nl7},
\arXiv{2505.12577}{gr-qc}.

\bibitem{Cheong:2025lwp}
S.~Cheong and W.~Kim,
``Strong gravitational lensing effects of black holes with quantum hair,''
\Journal{Phys. Rev. D}{\textbf{112}, 124041 (2025)}
{10.1103/8v7b-2lfs},
\arXiv{2508.07565}{gr-qc}.

\bibitem{Rodriguez:2025gfw}
B.~Rodr{\'\i}guez, I.~D{\'\i}az-Salda{\~n}a, W.~Yunpanqui and J.~Chagoya,
``Strong lensing by GUP-improved black holes,''
\Journal{Class. Quant. Grav.}{\textbf{43}, 035006 (2026)}
{10.1088/1361-6382/ae3afd},
\arXiv{2509.22880}{gr-qc}.

\bibitem{Kumar:2025mpb}
A.~Kumar and S.~G.~Ghosh,
``Strong gravitational lensing by black holes with a cloud of strings and constraints from EHT observations of M87* and Sgr A*,''
\Journal{Annals Phys.}{\textbf{484}, 170291 (2026)}
{10.1016/j.aop.2025.170291}.

\bibitem{Khodadi:2025upl}
M.~Khodadi, B.~Pourhassan and E.~N.~Saridakis,
``Multiprobe analysis of strong-field effects in f(Q) gravity,''
\Journal{Phys. Rev. D}{\textbf{113}, 064020 (2026)}
{10.1103/y2dx-7qgt},
\arXiv{2512.03529}{gr-qc}.

\bibitem{Weinberg:1972kfs}
S.~Weinberg,
``Gravitation and Cosmology: Principles and Applications of the General Theory of Relativity,''
John Wiley and Sons, 1972.

\bibitem{Gralla:2019xty}
S.~E.~Gralla, D.~E.~Holz and R.~M.~Wald,
``Black Hole Shadows, Photon Rings, and Lensing Rings,''
\Journal{Phys. Rev. D}{\textbf{100}, 024018 (2019)}
{10.1103/PhysRevD.100.024018},
\arXiv{1906.00873}{astro-ph.HE}.

\bibitem{Fauzi:2024nta}
M.~F.~Fauzi, H.~S.~Ramadhan and A.~Sulaksono,
``Anisotropic gravastar as horizonless regular black hole spacetime and its images illuminated by thin accretion disk,''
\Journal{Eur. Phys. J. C}{\textbf{84}, 1145 (2024)}{10.1140/epjc/s10052-024-13519-6}.

\bibitem{Zeng:2023fqy}
W.~Zeng, Y.~Ling, Q.~Q.~Jiang and G.~P.~Li,
``Accretion disk for regular black holes with sub-Planckian curvature,''
\Journal{Phys. Rev. D}{\textbf{108}, 104072 (2023)}
{10.1103/PhysRevD.108.104072},
\arXiv{2308.00976}{gr-qc}.

\bibitem{Meng:2023htc}
Y.~Meng, X.~M.~Kuang, X.~J.~Wang, B.~Wang and J.~P.~Wu,
``Images from disk and spherical accretions of hairy Schwarzschild black holes,''
\Journal{Phys. Rev. D}{\textbf{108}, 064013 (2023)}
{10.1103/PhysRevD.108.064013},
\arXiv{2306.10459}{gr-qc}.

\bibitem{Macedo:2024qky}
C.~F.~B.~Macedo, J.~L.~Rosa and D.~Rubiera-Garcia,
``Optical appearance of black holes surrounded by a dark matter halo,''
\Journal{JCAP}{\textbf{07}, 046 (2024)}
{10.1088/1475-7516/2024/07/046},
\arXiv{2402.13047}{gr-qc}.

\bibitem{Vagnozzi2023}
S.~Vagnozzi, R.~Roy, Y.~D.~Tsai, L.~Visinelli, M.~Afrin, A.~Allahyari, P.~Bambhaniya, D.~Dey, S.~G.~Ghosh and P.~S.~Joshi, \textit{et al.}
``Horizon-scale tests of gravity theories and fundamental physics from the Event Horizon Telescope image of Sagittarius A,''
\Journal{Class. Quant. Grav.}{\textbf{40}, 165007 (2023)}
{10.1088/1361-6382/acd97b},
\arXiv{2205.07787}{gr-qc}.

\bibitem{Gralla:2020srx}
S.~E.~Gralla, A.~Lupsasca and D.~P.~Marrone,
``The shape of the black hole photon ring: A precise test of strong-field general relativity,''
\Journal{Phys. Rev. D}{\textbf{102}, 124004 (2020)}
{10.1103/PhysRevD.102.124004},
\arXiv{2008.03879}{gr-qc}.

\bibitem{Guo:2022iiy}
S.~Guo, G.~R.~Li and E.~W.~Liang,
``Optical appearance of a thin-shell wormhole with a Hayward profile,''
\Journal{Eur. Phys. J. C}{\textbf{83}, 663 (2023)}
{10.1140/epjc/s10052-023-11842-y},
\arXiv{2210.03010}{gr-qc}.

\bibitem{Vincent:2022fwj}
F.~H.~Vincent, S.~E.~Gralla, A.~Lupsasca and M.~Wielgus,
``Images and photon ring signatures of thick disks around black holes,''
\Journal{Astron. Astrophys.}{\textbf{667}, A170 (2022)}
{10.1051/0004-6361/202244339},
\arXiv{2206.12066}{astro-ph.HE}.

\bibitem{Rosa:2023hfm}
J.~L.~Rosa,
``Observational properties of relativistic fluid spheres with thin accretion disks,''
\Journal{Phys. Rev. D}{\textbf{107}, 084048 (2023)}
{10.1103/PhysRevD.107.084048},
\arXiv{2302.11915}{gr-qc}.

\bibitem{Rosa:2024bqv}
J.~L.~Rosa, D.~S.~J.~Cordeiro, C.~F.~B.~Macedo and F.~S.~N.~Lobo,
``Observational imprints of gravastars from accretion disks and hot spots,''
\Journal{Phys. Rev. D}{\textbf{109}, 084002 (2024)}
{10.1103/PhysRevD.109.084002},
\arXiv{2401.07766}{gr-qc}.

\bibitem{daSilva:2023jxa}
L.~F.~D.~da Silva, F.~S.~N.~Lobo, G.~J.~Olmo and D.~Rubiera-Garcia,
``Photon rings as tests for alternative spherically symmetric geometries with thin accretion disks,''
\Journal{Phys. Rev. D}{\textbf{108}, 084055 (2023)}
{10.1103/PhysRevD.108.084055},
\arXiv{2307.06778}{gr-qc}.

\bibitem{Gao:2023mjb}
X.~J.~Gao, T.~T.~Sui, X.~X.~Zeng, Y.~S.~An and Y.~P.~Hu,
``Investigating shadow images and rings of the charged Horndeski black hole illuminated by various thin accretions,''
\Journal{Eur. Phys. J. C}{\textbf{83}, 1052 (2023)}
{10.1140/epjc/s10052-023-12231-1},
\arXiv{2311.11780}{gr-qc}.

\end{thebibliography}
\end{document}